\documentclass[aps,prb,twocolumn,superscriptaddress]{revtex4-2}

\usepackage{amsmath, amssymb,  graphicx, caption}

\usepackage[colorlinks=true ,urlcolor=blue,urlbordercolor={0 1 1}]{hyperref}

\usepackage{color}
\usepackage{xcolor}
\usepackage{braket}

\usepackage[utf8]{inputenc}
\usepackage{bm}
\usepackage{verbatim}
\usepackage{graphicx}
\usepackage{amsmath}
\usepackage{color}
\usepackage{float}
\usepackage{bbm}
\usepackage{amssymb}
\usepackage{slashed}
\usepackage{wasysym,bm,bbm,dsfont}

\begin{document}

\title{Fractional quantum anomalous Hall effects in  rhombohedral multilayer graphene in the moir\'eless limit}

\author{Boran Zhou}
\author{Hui Yang}
\author{Ya-Hui Zhang}
\affiliation{Department of Physics and Astronomy, Johns Hopkins University, Baltimore, Maryland 21218, USA}

\date{\today}

\begin{abstract}
 The standard theoretical framework for fractional quantum anomalous Hall effect (FQAH) assumes an isolated flat Chern band in the single particle level. In this paper we challenges this paradigm for the FQAH recently observed in the pentalayer rhombohedral stacked graphene aligned with hexagon boron nitride (hBN).  We show that  the external moir\'e superlattice potential  is simply a perturbation in a model with continuous translation symmetry. Through Hartree Fock calculation, we find that interaction opens a sizable remote band gap, resulting an isolated narrow $C=1$ Chern band at filling $\nu=1$. From exact diagonalization (ED) we identify FQAH phases at various fillings. But they exist also in the calculations without any external moir\'e potential. We suggest that the QAH insulator at $\nu=1$ should be viewed as an interaction driven topological Wigner crystal with QAH effect, which is then pinned by a small moir\'e potential. The $C=1$ QAH crystal is robust with a crystal period around $10\mathrm{nm}$  in 4-layer, 5-layer, 6-layer and 7-layer graphene systems. Our work suggests a new direction to exploring the interplay of topology and FQAH with spontaneous crystal formation in the vanishing moir\'e potential limit. We also propose a new system to generate and control both honeycomb and triangular moir\'e superlattice potential through Coulomb interaction from another control layer, which can stabilize or suppress the QAH crystal depending on the density of the control layer.  
\end{abstract}

\maketitle

\textbf{Introduction} There have been lots of efforts in realizing fractional quantum Hall (FQH) states\cite{stormer1999fractional,jain1989composite} from fractionally filling a narrow Chern band \cite{sun2011nearly,sheng2011fractional,neupert2011fractional,wang2011fractional,tang2011high,regnault2011fractional,bergholtz2013topological,parameswaran2013fractional,PhysRevB.84.165107} on a lattice.   Such a state is dubbed as fractional Chern insulator (FCI). FCI has been experimentally realized at non-zero magnetic field\cite{spanton2018observation,xie2021fractional}.  Fractional quantum anomalous Hall effect (FQAH) phase, a FCI at zero magnetic field, is more challenging. It was proposed that the two dimensional moir\'e systems are wonderful platforms to host nearly flat Chern band and thus FQAH states\cite{zhang2019nearly,ledwith2020fractional,repellin2020chern,abouelkomsan2020particle,wilhelm2021interplay} following spontaneous valley polarization\cite{zhang2019nearly,repellin2020ferromagnetism}. Indeed integer quantum anomalous Hall (QAH) states\cite{sharpe2019emergent,serlin2020intrinsic} were realized in twisted bilayer graphene (TBG) aligned with hexagon boron nitride (hBN)\cite{zhang2019twisted,bultinck2020mechanism}, in ABC stacked trilayer graphene moir\'e with hBN alignment\cite{chen2020tunable}, and also in transition Metal Dichalcogenide (TMD) bilayers\cite{li2021quantum,foutty2023mapping}.  More recently, FQAH was finally observed in twisted MoTe$_2$ homobilayer from optical and capacitance measurement\cite{cai2023signatures,zeng2023integer}, further supported directly from transport  measurements\cite{park2023observation,PhysRevX.13.031037}. Theoretically the existence of the FQAH (zero field FCI) in the twisted MoTe$_2$ system is quite natural due to the existence of isolated narrow Chern band\cite{wu2019topological,yu2020giant,devakul2021magic}. Indeed FQAH phases were predicted\cite{li2021spontaneous,crepel2023fci} even before the experiment. Recent theoretical works further confirm the existence of FQAH \cite{wang2023fractional,reddy2023fractional,2023arXiv230809697X,2023arXiv230914429Y} and also composite Fermi liquid (CFL)\cite{PhysRevLett.131.136501,PhysRevLett.131.136502}  at even denominator. Apparently twisted MoTe$_2$ system mimics the familiar lowest Landau level physics quite well.

Possibility of FQAH was also discussed in graphene systems\cite{zhang2019nearly,ledwith2022family,wang2022hierarchy,devakul2023magic,wang2023origin,gao2023untwisting,ghorashi2023topological}, but so far it has been reported only in  pentalayer ($5$-layer) rhombohedral stacked graphene aligned with hBN\cite{2023arXiv230917436L}, which is actually a surprise and unexpected theoretically. In the following we always refer $n$-layer graphene to rhombohedral stacked multilayer graphene. It was predicted by one of us\cite{zhang2019nearly,zhang2019bridging} that there is a $C\neq 0$ band in  $n$-layer graphene aligned with hBN for one sign of the displacement field $D$, corresponding to the side where particles are pushed away from the hBN. See also Ref.~\cite{PhysRevB.87.241108}  for earlier discussions on possible narrow bands with non-zero Berry curvature in the system. The superlattice potential from the hBN alignment on top only applies on the top layer of the $n$-layer graphene. Therefore, for the topological side of $D$, electrons feel a weaker superlattice potential and the remote gap tends to be small.  This effect is more severe for larger $n_{\mathrm{layer}}$ because $D$ can polarize the layer degree of freedom more easily. For the $5$-layer graphene, FQAH is found only in the strong displacement region and we estimate that the moire superlattice potential projected to the conduction band is only at order of $0.05\mathrm{meV}$. Such a small superlattice potential is only a perturbation term and we indeed notice that the band structures with and without moir\'e potential are roughly the same.  Hence it is a surprise that even QAH insulator can be stabilized at $\nu=1$ filling per moir\'e unit cell.

We resolve this puzzle by performing a Hartree Fock (HF) calculation at filling $\nu=1$ first. Assuming spin-valley polarization\cite{repellin2020ferromagnetism}, we find a $C=1$ Chern insualtor in a range of displacement field $D$ at twist angle $\theta \in (0.70^\circ,1.40^\circ)$ with sizable band gap and narrow bandwidth of the filled band. Then at fractionally filling of the HF renormalized band, we find FQAH insulators through exact diagonalization (ED). There is also signature of composite Fermi liquid\cite{halperin1993theory} at 1/2 filling.  In this calculation, the external moir\'e potential does not play any essential role. The same result can be reproduced by a calculation with the moir\'e potential turned off by hand. In this sense, we should view the $\nu=1$ Chern insulator as a QAH crystal spontaneously breaking  the approximate continuous translation symmetry, which is then pinned by the small external moir\'e potential. The physics is similar to the Hall crystal proposed at high magnetic field\cite{tevsanovic1989hall}. In our picture, the FQAH is then realized at fractional filling of this spontaneously formed crystal. Our work thus establishes the n-layer graphene as a completely new platform for FQAH phase, distinct from twisted MoTe$_2$. We note that the picture of viewing moir\'e as a perturbation has already been suggested in Ref.\onlinecite{zhou2021half} in the trilayer graphene system, though in that case there is still controversy and the main focus is on the valence band\cite{patri2023strong}. When the $n_{\mathrm{layer}}\geq 4$, the conduction band is more flat and the projected moir\'e potential is clearly small, hence the physics is closer to the moir\'eless limit with approximate continuous translation symmetry. From HF calculation, we find Chern insulator with narrow $C=1$ band at $\nu=1$ in a range of the parameter space $(a_M, D)$ with $a_M\approx 10$ nm for $n_{\mathrm{layer}}=4,5,6,7$. Thus we propose to search for QAH and FQAH phases also with $n_{\mathrm{layer}}=4,6,7$.

One interesting question is whether it is possible to realize the QAH crystal in the moir\'eless n-layer graphene\cite{han2023correlated,han2023large,han2023orbital,liu2023interaction}. In our HF calculation, the external moir\'e potential does not play any essential role. However, it is known that Hatree-Fock overestimates the strength of the insulator. This is because we restrict to only Slater Determinant states and the QAH insulator always wins over the Fermi liquid state without crystal formation. However, the Fermi liquid we can access is only the free fermion state and we can not rule out that a more complicated wavefunction of a correlated metal actually has better energy at zero moir\'e potential.  On the other hand, in the strong moir\'e potential limit, an insulator at the filling $\nu=1$ is quite natural and the interaction driven crystal can be stabilized by the external moir\'e potential. With the above arguments, there is a possibility that the QAH insulator ansatz found by our HF calculation is the true ground state only when the external moir\'e potential is above a small threshold $V_M^c$.  To check this scenario, we propose a new system with the moir\'e potential generated by Coulomb interaction from a control layer such as TBG. The moir\'e potential can then be tuned continuously to obtain the critical threshold $V_M^c$ experimentally.

\textbf{Model} We model the $n$-layer graphene aligned with hexagonal boron nitride (hBN) as $H_K=H_0+H_M$, where $H_0$ is the Hamiltonian of $n$-layer graphene and $H_M$ is the effective moir\'e potential from hBN alignment. It is by now well established that the $\nu=1$ QAH insulator is spin-valley polarized from the Coulomb exchange\cite{repellin2020ferromagnetism}. The real puzzle is how to obtain a Chern band within one spin-valley flavor. Let us focus on the valley $K$ and define $\psi_z(\mathbf k)=(f_{z;A}(\mathbf k),f _{z;B}(\mathbf k))^T$ for the two sublattice at the layer $z$. The Hamiltonian for the other valley $K'$ is related by time reversal symmetry. The free Hamiltonian for the valley $K$ is written down as a $2n_\mathrm{layer}$ band model:

\onecolumngrid

\begin{align}
    H_0=&\sum_{\mathbf k}\sum_{z=1}^{n_\mathrm{layer}} \psi^\dagger_z(\mathbf k)\begin{pmatrix} V_{z,l}+u_{A,l}& -v_1 (k_x-\mathrm{i}k_y)e^{\mathrm{i}\theta_3} \\ -v_1(k_x+\mathrm{i}k_y)e^{-\mathrm{i}\theta_3} &V_{z,l}+u_{B,l}\end{pmatrix}\psi_z(\mathbf k) +\sum_{\mathbf k}\sum_{z=1}^{n_\mathrm{layer}-2}\psi_z^\dag({\bf k})\begin{pmatrix} 0 & \frac{\gamma_2}{2} \\ 0 &0\end{pmatrix}\psi_{z+2}({\bf k})+\mathrm{H.c}.
    \label{eq:free_full_H}\notag \\
    &+\sum_{\mathbf k}\sum_{z=1}^{n_\mathrm{layer}-1}\psi_z^\dag({\bf k})\begin{pmatrix} -v_4(k_x-\mathrm{i}k_y)e^{\mathrm{i}\theta_3} & -v_3 (k_x+\mathrm{i}k_y)e^{-\mathrm{i}\theta_3} \\ \gamma_1 &-v_4(k_x-\mathrm{i}k_y)e^{\mathrm{i}\theta_3}\end{pmatrix}\psi_{z+1}({\bf k})+\mathrm{H.c.}\notag\\
\end{align}

\twocolumngrid
where the parameters $v_i=\frac{\sqrt{3}}{2}t_i$, with \cite{2023PhRvB.108o5406P,PhysRevB.82.035409} $t_1=-2600\mathrm{meV}$, $t_3=293\mathrm{meV}$, $t_4=144\mathrm{meV}$, $\gamma_1=358\mathrm{meV}$, $\gamma_2=-8.3\mathrm{meV}$ correspond to the hoppings. $V_{z,l}$ is the layer-dependent potential from the displacement field with $V_{z,l}=\frac{D(l-1-n_\mathrm{layer}/2)}{n_\mathrm{layer}-1}$, where $D$ is the displacement field, and $l=1,2,\cdots$ is the $l$-th layer. $u_{A,1}=u_{B,n_\mathrm{layer}}=0, u_{B,1}=u_{A,n_\mathrm{layer}}=12.2\mathrm{meV}, u_{A,l}=u_{B,l}=-16.4\mathrm{meV}$ for other terms. The phase $\theta_3$ is introduce to rotate the graphene. We have $\theta_3=\arctan\frac{\theta}{\delta}$ with $\theta$ as the twist angle between graphene and hBN. $\delta=\frac{a_\mathrm{hBN}-a_\mathrm{G}}{a_\mathrm{hBN}}=0.017$,  $a_\mathrm{G}$(0.246nm) and $a_\mathrm{hBN}$(0.25025nm) are the lattice constant of graphene and hBN respectively.

 \begin{figure}[tbp]    
    \includegraphics[width=0.9\linewidth]{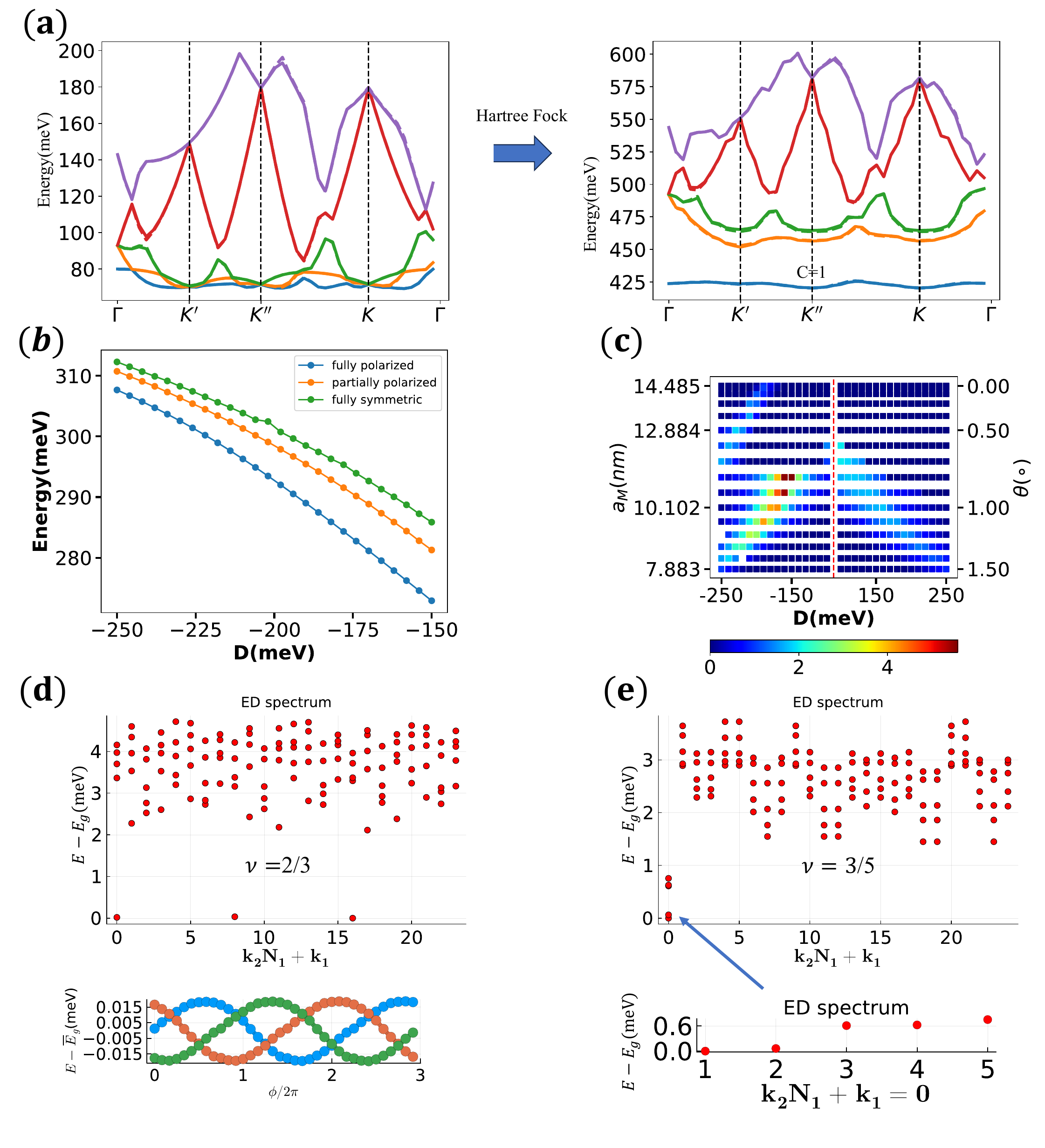}
    \caption{Numerical results of the 5-layer graphene/hBN system at $\theta=0.77^\circ(a_M=11.36\mathrm{nm})$, $\epsilon=6$, $D=-160\mathrm{meV}$. (a) Conduction band structure before and after performing HF calculation at $\nu=1$ using $24\times24$ points in the moir\'e Brillouin zone (MBZ) and keeping $N_b=5$ bands.  The solid line and the dashed line corresponds to $V_M=1,0$ respectively.  They are basically identical to each other. (b) The  energy per unit cell at $\nu=1$ of three different ansatz: fully spin-valley polarized, valley polarized and spin symmetric (partially polarized), fully symmetric. (c) $V_M=1$, dependence of $\frac{|C|\Delta}{W}$ on $D$ and $a_M$. HF calclualtion is performed in $12\times18$ system. (d) ED calculation at $\nu=2/3$ filling with a system size of $4\times 6$.  $k_1$ and $k_2$ is defined as $\mathbf{k}=k_1 \mathbf{G_1}/N_1+k_2\mathbf{G_2}/N_2$ with the system size of $N_1\times N_2$. We also show the spectral flow under flux insertion along $k_2$ direction.  Clearly there is a permutation of the degenerate states under a flux of $\phi=2\pi$. (e) ED calculation at $\nu=3/5$ filling with system size $5\times 5$.  In the lower part we show a zoom in of  the five nearly degenerate states.  In this paper, we use $N_b=5$ unless specified.}
    \label{fig:FCI_evidence}
\end{figure}

 The moir\'e potential term is $H_M=\sum_{\mathbf{k}}\sum_{j=1}^6\psi^\dag_1(\mathbf{k})H_M(\mathbf{G_j})\psi_1(\mathbf{k+G_j})$
which only acts on the first layer aligned with the hBN. Here the moir\'e reciprocal lattice vectors are $\mathbf{G_j}=\frac{4\pi}{\sqrt{3}a_M}(\cos(\frac{j\pi}{3}-\frac{5\pi}{6}),\sin(\frac{j\pi}{3}-\frac{5\pi}{6}))^T,j=1,2\cdots,6$. The moir\'e potential parameters are listed \cite{PhysRevB.89.205414,PhysRevB.82.035409,2023PhRvB.108o5406P} in the supplementary.  In the following we will also introduce a parameter $V_M$ to scale the moir\'e potential term by $H_M \rightarrow V_M H_M$. We mainly focus on $V_M=1,0$.

In addition to the kinetic energy, we also have the Coulomb interaction:

\begin{equation}
    H_V=\frac{1}{2A} \sum_{l,l'}\sum_{\mathbf q} V_{l l'}(q):\rho_l(q) \rho_{l'}(-q):
\end{equation}
where $A$ is the area of the sample, $\rho_l(q)$ is the density operator in the layer $l$. $V_{ll'}(\mathbf q)= \frac{e^2\tanh q \lambda}{2\epsilon_0 \epsilon q} e^{- q |l-l'|d_\mathrm{layer}}$, $\lambda=30\mathrm{nm}$ is the screening length and $d_\mathrm{layer}=0.34\mathrm{nm}$ is the distance between
nearest two layers. 

\textbf{Numerical evidence of QAH and FQAH} We focus on the conduction band on the large and negative $D$, so electrons in the conduction band are mainly staying in the bottom, far away from the aligned hBN on the top. In Fig.~\ref{fig:FCI_evidence}(a) we can clearly see that there is no remote band gap for the 5-layer graphene at $\theta=0.77^\circ$, $D=-160$ meV. Actually the band structure is basically the same as the band of the moir\'eless 5-layer graphene folded into the moir\'e Brillouin zone (MBZ). Then with a HF calculation (see the supplementary) keeping the lowest $N_b=5$ number of  conduction bands at the filling $\nu=1$ per moir\'e unit cell, we find a sizable band gap opening and a filled nearly flat $C=1$ band (see Fig.~\ref{fig:FCI_evidence}(a)).  This indicates that the $\nu=1$ filling is  an interaction driven QAH insulator. Note that in our HF calculation the valence bands are assumed to be fully filled due to the large band gap at large $|D|$. As the density from the valence band is uniform, the valence bands can be safely ignored. Here our ansatz is fully spin-valley polarized. To confirm the spin-valley ferromagnetism, we plot the HF energies for spin-valley polarized, partially polarized and fully symmetric ansatz in Fig.~\ref{fig:FCI_evidence}(b), it is clear that the fully spin-valley polarized state always has the lowest energy.  In Fig.~\ref{fig:FCI_evidence}(c) we show the color map of $\frac{|C| \Delta}{W}$, with $\Delta$ and $W$ as the band gap and bandwidth of the first band from the HF calculation. The brightest region is along a stripe with the moir\'e period around $10$ nm and $D$ negative. In  this region we have a $C=1$ band with large $\Delta$ and small $W$, ideal for FCI.

Then we perform ED projected to the HF renormalized lowest band. We  use the hole picture with HF renormalized dispersion relative to the fully filled band insulator at $\nu=1$.  This also means that the spin-valley is polarized as inherited from the parent state at $\nu=1$. In Fig.~\ref{fig:FCI_evidence}(d)(e) we can see low lying 3-fold and 5-fold states, consistent with FCI at electron filling $\nu=2/3$ and $\nu=3/5$ respectively.  At $\nu=\frac{1}{2}$, we also find signatures of composite Fermi liquid (CFL) (see the supplementary).

\begin{figure}[tbp]    
    \includegraphics[width=0.9\linewidth]{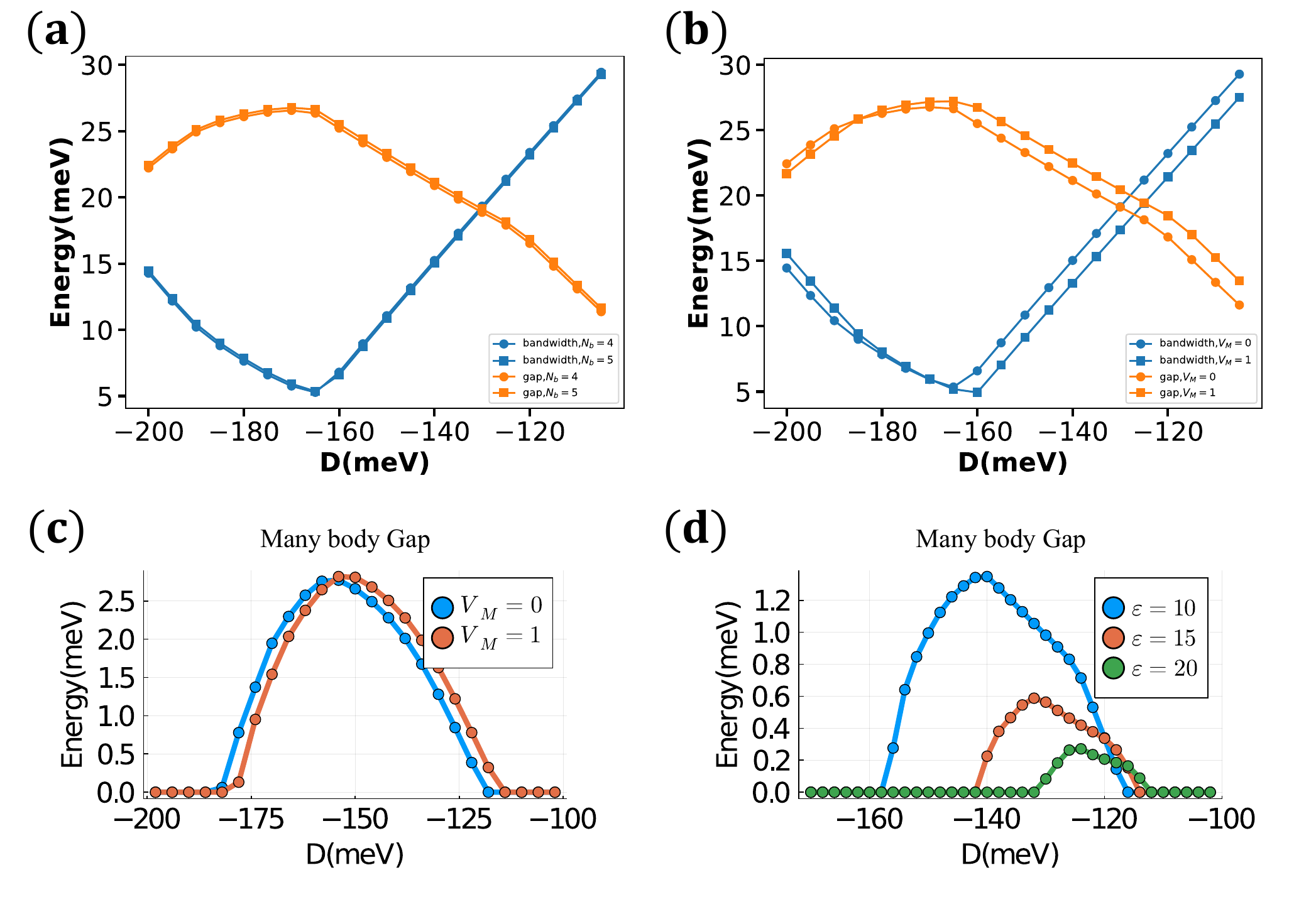}
    \caption{$\theta=0.9^\circ(a_M=10.63\mathrm{nm})$. (a)(b) $\epsilon=6$, dependence of $\Delta$ and $W$ on $D$. (a) The results are the same for the number of kept bands $N_b=4,5$. Here we use $V_M=0$. (b) $V_M=0,1$ give roughly the same results, confirming that the external moir\'e potential has very weak effect in determing the band properties. (c) $\epsilon=6$, dependence of the many body gap of the FQAH insulator at $\nu=2/3$ filling on $D$. The many body gap is defined as the energy difference between the fourth and the third states in the ED calculation. The ED calculation is performed by using $4\times6$ points in MBZ.  (d) Similar to (c), but with different value of $\epsilon$ and $V_M=0$. }
    \label{fig:phase_diagram}
\end{figure}


\begin{figure}[tbp]    
    \includegraphics[width=0.9\linewidth]{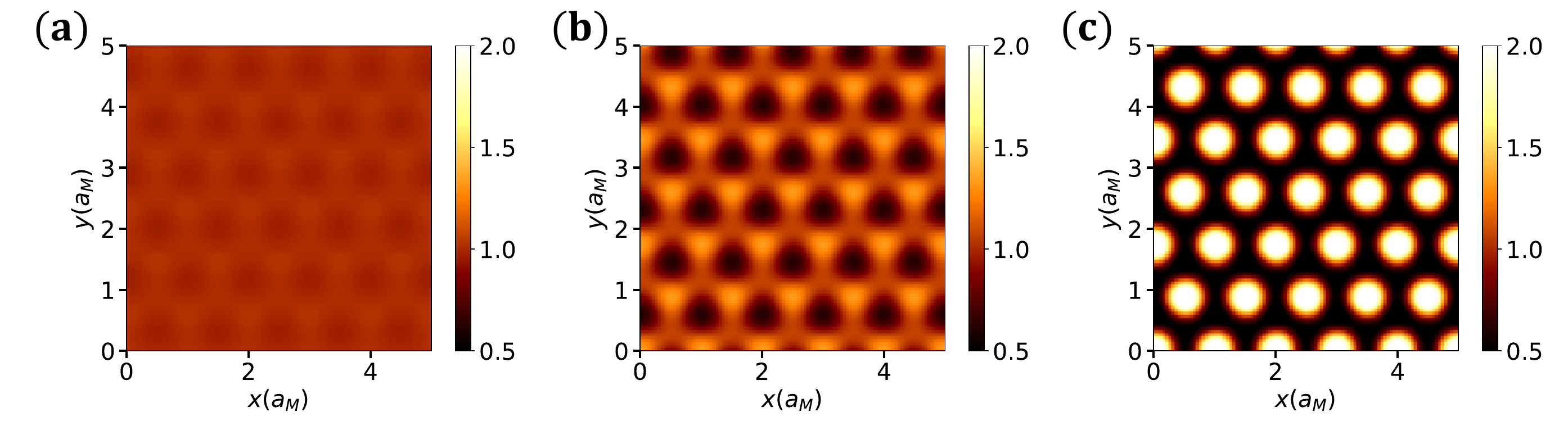}
    \caption{The density distribution $\rho(\mathbf{r})$ at $\nu=1,\epsilon=6,D=-160\mathrm{meV}$ of the 5-layer graphene/hBN system: (a) $\theta=0.77^\circ$, without HF; (b) $\theta=0.77^\circ$, with HF, the lowest HF band has $C=1$; (c) $\theta=0.3^\circ$, with HF, the lowest HF band has $C=0$. }
    \label{fig:density_profile}
\end{figure}

\begin{figure}[tbp]    
    \includegraphics[width=0.9\linewidth]{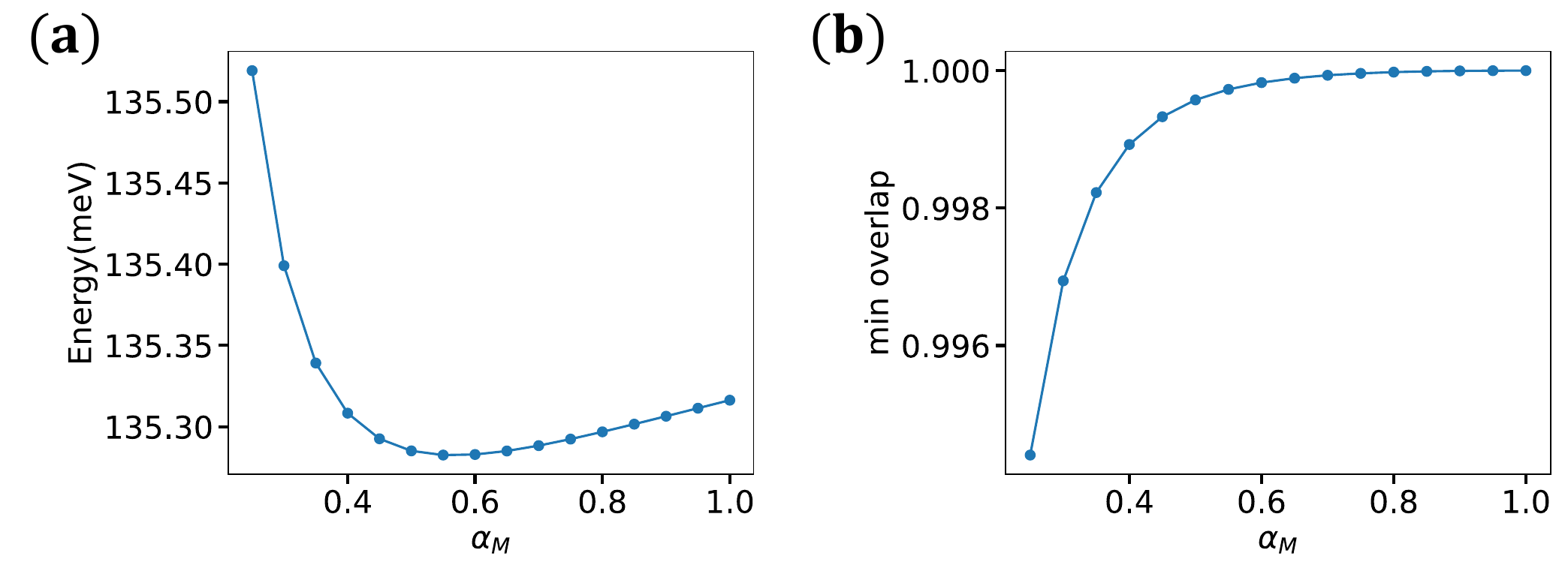}
    \caption{Improved ED calculation at $\nu=\frac{2}{3}$ of the 5-layer graphene/hBN system at $\theta=0.77^\circ(a_M=11.36\mathrm{nm})$, $\epsilon=6$, $D=-160\mathrm{meV}$. The system size is the same as Fig.~\ref{fig:FCI_evidence}. We keep $N_b=4$ conduction bands in our calculation. (a) Dependence of the energy per moir\'e unit cell on the variational parameter $\alpha$ at $\nu=\frac{2}{3}$. (b) Dependence of the minimal overlap of the deformed band with the band of $H_{\alpha=1}$ at each $\alpha$.}    \label{fig:change_alphaM_compare_FCI_AHC}
\end{figure}

\textbf{Phase diagram} 
In Fig.~\ref{fig:phase_diagram} we show the results along a line cut with varying $D$ at fixed $\theta=0.9^\circ$. First, we demonstrate that the calculation is well converged using $N_b=5$ in Fig.~\ref{fig:phase_diagram}(a). In a narrow range of displacement field around $D=-160$ meV, $W$ becomes small and $\Delta$ becomes large (see Fig.~\ref{fig:phase_diagram}(b)). This corresponds to exactly the region that the many body gap of the FCI state at $\nu=\frac{2}{3}$ is maximized from our ED calculation (see Fig.~\ref{fig:phase_diagram}(c)).  The FCI state is also stable to the renormalized dielectric constant $\epsilon$ which controls the interaction strength. From our calculation, the FCI gap is finite even when $\epsilon=20$ (see Fig.~\ref{fig:phase_diagram}(d)). We also find that the same $C=1$  QAH crystal at $\nu=1$ exists also for $n_{\mathrm{layer}}=4,6,7$ as shown in the supplementary.


\textbf{The role of the moir\'e potential}  Our result in Fig.~\ref{fig:phase_diagram} clearly shows that the external moir\'e potential term is not essential for QAH and FQAH phases in this system. We project the moir\'e potential into the conduction band, the matrix element is around $0.03\mathrm{meV}$ and $0.05\mathrm{meV}$ at $K, K^\prime$ points in the MBZ respectively, which is obviously just a tiny perturbation. Therefore, we conclude that the narrow Chern band shown in Fig.~\ref{fig:FCI_evidence}(a) is completely from the Coulomb interaction. Our $\nu=1$ QAH insulator survives to the $V_M \rightarrow 0$ limit and should be viewed as a crystal spontaneously breaking the continuous translation symmetry. We plot the density profile $\langle \rho(\mathbf r) \rangle$ for the $\nu=1$ state in Fig.~\ref{fig:density_profile}. The density is basically uniform without the HF generated potential (see Fig.~\ref{fig:density_profile}(a)). The HF calculation generates a crystal with  $C=1$ or $C=0$  at different twist angles. The $C=0$ ansatz is just an usual triangular lattice Wigner crystal, while the QAH crystal has a honeycomb lattice structure with finite but not full sublattice polarization (see Fig.~\ref{fig:density_profile}(b)). In this picture, the FQAH phases are then realized at fractional filling of this QAH crystal.

\begin{figure}[tbp]    
    \includegraphics[width=0.95\linewidth]{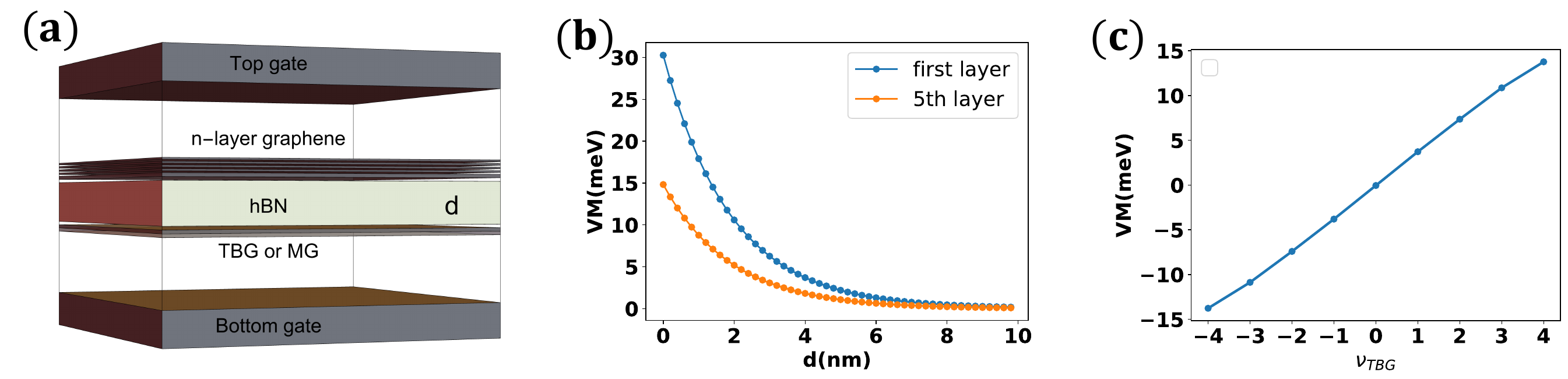}
    \caption{(a)Illustration of the $n$-layer graphene/hBN/TBG setup with hBN misaligned with both top and bottom graphene. The Coulomb interaction between the TBG and the $n$-layer graphene will effectively generate a moir\'e potential in the $n$-layer graphene. We fix the density of TBG in the fully filled filling $\nu_\mathrm{TBG}=4$. The TBG can also be replaced with a monolayer graphene (MG) aligned with the middle hBN. (b) $d$-dependence of the moir\'e potential $V_M$ in the first layer and 5th-layer graphene. The TBG is at the magic angle $\theta=1.02^\circ$ with dielectric constant $\epsilon=6$. (c) filling-dependence of the moir\'e potential $V_M$ in the remote layer at $d=1.5\mathrm{nm}$. Here $\nu_{\mathrm{TBG}}$ is the filling of the TBG layer.}
    \label{fig:setup}
\end{figure}

\textbf{Discussions on the fractional filling} At fractional filling, our ED calculation above assumeed a rigid band from the HF calculation at $\nu=1$. To justify this approximation, we also perform an improved variational calculation. First let us assume a same crystal period $a_{\text{crystal}}=a_M$ also for the fractional filling. We perform the ED calculation by projecting the original Hamiltonian $H=H_K+H_V$ to the first band of a variational mean field Hamiltonian $H_\alpha=H_K+\alpha H_\mathrm{HF}$ with $\alpha$ as a parameter to be optimized.  $H_\mathrm{HF}$ is the additional Hartree and Fock terms from the HF calculation at $\nu=1$.  In the ED calculation we use the original Hamiltonian $H=H_K+H_V$ to obtain the ground state $\ket{\psi_{\alpha}}$ and energy $E(\alpha)$ (see the supplementary). At $\nu=\frac{2}{3}$ the optimal $\alpha$ minimizing $E_{\alpha}$ is $\alpha \approx 0.5$, as shown in Fig.~\ref{fig:change_alphaM_compare_FCI_AHC}(a). So the interaction driven crystal is not destroyed. Actually we find that the  band at this value of $\alpha=0.5$ has an overlap of more than $0.99$ at every momentum with the band at $\nu=1$ ($\alpha=1$), as shown in Fig.~\ref{fig:change_alphaM_compare_FCI_AHC}(b). Therefore the projected band and thus the ED spectrum is only slightly modified compared to the calculation before (see the supplementary). We conclude that the rigid band approximation is valid and the FCI state is a stable ansatz.

Our picture of a topological Wigner crystal at the $V_M=0$ limit actually implies two possible competing states at the fractional filling such as $\nu=\frac{2}{3}$: (I) A FQAH state with the crystal pinned to the same lattice size as the moir\'e lattice constant $a_M$, as discussed above; (II) An extended integer QAH state with the spontaneously generated Wigner crystal period $a_{\text{crystal}}$ locked to the filling $\nu$ as $a_{\text{crystal}}=\frac{a_M}{\sqrt{\nu}}$. In addition to the FQAH state in Ref.~\cite{2023arXiv230917436L}, future experiments may find an integer QAH crystal phase instead in certain parameter regimes.

\textbf{Coulomb induced moir\'e}  To better study the effect a finite moir\'e potential,  it is desirable to control the external moir\'e potential experimentally. We now propose a different route to generate superlattice potential in the $n$-layer graphene. As is illustrated in Fig.~\ref{fig:setup}(a), we have $n$-layer graphene separated with a TBG by a thin and misaligned hBN with thickness $d$. TBG has moir\'e lattice constant $a_M=a_\mathrm{G}/(2\sin\frac{\theta}{2})$ controlled by its twist angle $\theta$. The Coulomb interaction between the TBG layer and the $n$-layer graphene will effectively induce a moir\'e potential in the $n$-layer graphene layers. The TBG has a charge profile depending on its filling $\nu_\mathrm{TBG}$, which generates superlattice potential felt by  each layer of the $n$-layer graphene: 
$V_{l}({\bf G}_i)=\frac{2}{\sqrt{3}}\frac{e^2}{2\epsilon\epsilon_0a_M}e^{-G_M{(d+ld_\mathrm{layer})}}\frac{1}{G_Ma_M}\langle \rho({\bf G}_M) \rangle$, where $\langle \rho({\bf G}_M) \rangle$ is the Fourier transformation of  the density profile in the TBG and $d_\mathrm{layer}=0.34\mathrm{nm}$ is the distance between the adjacent graphene layers. $l=0,1,...$ is the layer index count from the one closest to the TBG. As is shown in Fig.~\ref{fig:setup}(b), the moir\'e potentials can be generated in both the adjacent layer and the remote layer and are controlled by the hBN thickness $d$.  The moir\'e potential can also be tuned by the density in the TBG layer (see Fig.~\ref{fig:setup}(c)). Especially the $\nu_{\mathrm{TBG}}=4$ of TBG generates a honeycomb lattice potential, while $\nu_{\mathrm{TBG}}=-4$ generates a triangular lattice. The QAH crystal is weaken by the triangular lattice potential, while it is stable under the honeycomb lattice potential (see the supplementary).

\textbf{Summary} In conclusion, we studied the recently observed FQAH states in the pentalayer rhombohedral stacked graphene aligned with hBN and in related systems.  In contrast to the standard framework for other moir\'e systems, we find that the external moir\'e potential is negligible and the QAH and FQAH phases should be understood as from a spontaneously formed  topological crystal due to interaction in a model with continuous translation symmetry. The same QAH crystal is also robust in 4-layer, 6-layer and 7-layer systems. In future it is interesting to study the interplay between the almost gapless phonon modes of the crystal with the small moir\'e potential and disorder.  At fractional filling such as $\nu=\frac{2}{3}$, our theory implies the possible competitions between (I) an integer QAH crystal with $\sigma_{xy}=\frac{e^2}{h}$ over a continuous range of dopings and (II) FQAH states. Besides, we propose a new setup with moir\'e in the $n$-layer graphene generated from the Coulomb repulsion of twist bilayer graphene (TBG) separated by a thin hBN. The ability to control the moir\'e potential can reveal more information on whether a small moir\'e potential is necessary to stabilize the QAH crystal and FQAH states. It is also interesting to dope the TBG layer, which screens the Coulomb in the n-layer graphene and may drive a transition out of the quantum Hall phases\cite{song2023phase}.

\textit{Note added}: We note two other  papers\cite{dong2023theory,dong2023anomalous} on the same topic also appeared roughly at the same time. Our results agree with each other when they overlap.  In the updated version (July 2024), we added an improved variational ED calculation to allow the band to deform with the doping. When we finalized the updated version, we also became aware of anothe preprint\cite{huang2024self} which also performed improved ED calculations beyond the rigid band assumption.

\textbf{Acknowledgement} YHZ thanks Zhihuan Dong and T. Senthil for many inspiring discussions and previous collaborations on moir\'e systems. We especially thank Zhihuan Dong for helps on improving the efficiency of the  Hartree Fock calculation.  We also thank Long Ju and Sankar Das Sarma for discussions and useful comments. The numerical simulation was  carried out at the
Advanced Research Computing at Hopkins (ARCH) core facility (rockfish.jhu.edu), which is supported by the National
Science Foundation (NSF) grant number OAC 1920103. This work was supported by the National Science Foundation under Grant No. DMR-2237031.

\bibliographystyle{apsrev4-1}
\bibliography{main}

\appendix

\widetext \vspace{0.5cm}

\section{The moir\'e potential form}

The moir\'e potential parameters used in our calculation are listed in the following:

\begin{align}
    H_z(\mathbf {G_1})&=H_z(\mathbf{G_3})=H_z(\mathbf{G_3})=C_{z}e^{\mathrm{i}\phi_z}\notag\\
    &=H_z^*(\mathbf{G_2})=H_z^*(\mathbf{G_4})=H_z^*(\mathbf{G_6}),\\
    H_0(\mathbf{G_1})&=H_0(\mathbf{G_3})=H_0(\mathbf{G_5})=C_{0}e^{\mathrm{i}\phi_0}\notag\\
    &=H_0^*(\mathbf{G_2})=H_0^*(\mathbf{G_4})=H_0^*(\mathbf{G_6}),\\
    H_{AB}(\mathbf{G_1})&=H_{AB}^*(\mathbf{G_4})=C_{AB}e^{\mathrm{i}(2\pi/3-\phi_{AB})},\\
    H_{AB}(\mathbf{G_3})&=H_{AB}^*(\mathbf{G_2})=C_{AB}e^{-\mathrm{i}\phi_{AB}},\\
    H_{AB}(\mathbf{G_5})&=H_{AB}^*(\mathbf{G_6})=C_{AB}e^{\mathrm{i}(-2\pi/3-\phi_{AB})},
\end{align}
where $C_{0}=-10.13\mathrm{meV}$, $\phi_0=86.53^\circ$, $C_z=-9.01\mathrm{meV}$, $\phi_z=8.43^\circ$, $C_{AB}=11.34\mathrm{meV}$, $\phi_{AB}=19.60^\circ$.  Here $H_{AA}(\mathbf{G_j})=H_{0}(\mathbf{G_j})+H_z(\mathbf{G_j})$ and $H_{BB}(\mathbf{G_j})=H_0(\mathbf{G_j})-H_z(\mathbf{G_j})$ for each $\mathbf {G_j}$.

\section{Hartree Fock calculation}

We perform Hartree Fock approximation by keeping the first $N_b$ conduction bands in the MBZ. The interaction term is written as:
\begin{equation}\label{EqInteraction}
    H_{V}=\frac{1}{2A}\sum_\mathbf{q} \sum_{l,l^\prime} V_{l,l^\prime}(\mathbf{q}) :\rho_l(\mathbf{q})\rho_{l^\prime}(\mathbf{-q}):,
\end{equation}
where $\rho_l(\mathbf{q})=\sum_{\mathbf{k},m,m^\prime}c^\dagger_m(\mathbf{k+q})\Lambda^l_{m,m^\prime}(\mathbf{k},\mathbf{q})c_{m^\prime}(\mathbf{k})$ and $\Lambda^l_{m,m^\prime}(\mathbf{k},\mathbf{q})=\langle u_m({\mathbf{k+q}})|P_l|u_{m^\prime}(\mathbf{k})\rangle$, $P_l$ is the projection operator to layer $l$, $m$ and $m^\prime$ are band indices. $V_{l,l^\prime}(\mathbf{q})=\frac{e^2e^{-q|l-l^\prime|d_{\mathrm{layer}}}\tanh{(q\lambda)}}{2\epsilon\epsilon_0|q|}$, where $d_{\mathrm{layer}}$ is the distance between nearest two layers, $\lambda$ is the screening length. In our calculation, we choose that $d_\mathrm{layer}=0.34\mathrm{nm}$ and $\lambda=30\mathrm{nm}$. The interaction can be decoupled into Hartree and Fock terms respectively, leading to a mean field theory:
\begin{equation}
\begin{split}
H_{V}=&\sum_{\mathbf{k_1},\mathbf{k_2}}\sum_{\mathbf{q}}\sum_{m,m^\prime,n,n^\prime}V_{m,m^\prime,n,n^\prime}(\mathbf{k_1},\mathbf{k_2},\mathbf{q})c^\dagger_m(\mathbf{k_1+q})c^\dagger_n(\mathbf{k_2-q})c_{n^\prime}(\mathbf{k_2})c_{m^\prime}(\mathbf{k_1})\\
=&\sum_{\mathbf{k_1},\mathbf{k_2}}\sum_{m,m^\prime,n,n^\prime}2\left(V_{m,m^\prime,n,n^\prime}(\mathbf{k_1},\mathbf{k_2},\mathbf{0})-V_{n,m^\prime,m,n^\prime}(\mathbf{k_1},\mathbf{k_2},\mathbf{k_2-k_1})\right)\langle c^\dagger_m(\mathbf{k_1})c_{m^\prime}(\mathbf{k_1})\rangle c_n^\dagger(\mathbf{k_2})c_{n^\prime}(\mathbf{k_2}).
\end{split}
\end{equation}
The interaction vertex $V_{m,m^\prime,n,n^\prime}(\mathbf{k_1},\mathbf{k_2},\mathbf{q})$ can be calculated as:
\begin{equation}
V_{m,m^\prime,n,n^\prime}(\mathbf{k_1},\mathbf{k_2},\mathbf{q})=\frac{1}{2A}\sum_{l,l^\prime}V_{l,l^\prime}(\mathbf{q})\Lambda^l_{m,m^\prime}(\mathbf{k_1},\mathbf{q})\Lambda^l_{n,n^\prime}(\mathbf{k_2},\mathbf{-q}).
\end{equation}

In the above $m,m',n,n'=1,2,...,N_b$ are the band indexes. In the calculation we solve $\rho_{mm'}(\mathbf k)=\langle c^\dagger_m(\mathbf k) c_{m'}(\mathbf k) \rangle$ self consistently from randomized initial ansatz. We try $40$ number of randomized initial ansatz and choose the one with lowest energy. In this HF calculation we use the charge neutrality scheme, in the sense that the reference state is simply to occupy all of the valence band states.

\section{ED calculation}\label{Section_ED_calculation}
We perform the ED calculation by projecting the Coulomb interaction into the lowest HF band. The total Hamiltonian is given by:
\begin{equation}
    H=\sum_\mathbf{k} \epsilon(\mathbf{k})\tilde{c}_0^\dagger(\mathbf{k})\tilde{c}_0(\mathbf{k})+\sum_{\mathbf{k_1},\mathbf{k_2},\mathbf{q}}V(\mathbf{k_1},\mathbf{k_2},\mathbf{q})\tilde{c}_0^\dagger(\mathbf{k_1+q})\tilde{c}_0^\dagger(\mathbf{k_2-q})\tilde{c}_0(\mathbf{k_2})\tilde{c}_0(\mathbf{k_1}),
\end{equation}
where $\tilde{c}_{0}^\dagger(\mathbf{k})=U_{0n}(\mathbf{k})c_n^\dagger(\mathbf{k})$ is the electron operator of the lowest HF band, $c_n^\dagger(\mathbf{k})$ is the electron operator in the band of the bare $H_K$, $n$ is the band index. $\epsilon(\mathbf{k})=\sum_n |U_{0n}(\mathbf{k})|^2\epsilon_n(\mathbf{k})$ is the dispersion of the bare kinetic term $H_K=H_0+V_M H_M$ projecting onto the lowest HF band, $\epsilon_n(\mathbf{k})$ is the $n^\mathrm{th}$ band dispersion of $H_K$. The interaction vertex is in the form:
\begin{equation}\label{eqVertex}
    V(\mathbf{k_1},\mathbf{k_2},\mathbf{q})=\frac{1}{2A}\sum_{l,l^\prime} V_{l,l^\prime}(\mathbf{q})\Lambda^l_\mathrm{HF}(\mathbf{k_1},\mathbf{q})\Lambda^{l^\prime}_\mathrm{HF}(\mathbf{k_2},\mathbf{-q}),
\end{equation}
where $\Lambda^l_\mathrm{HF}(\mathbf{k_1},\mathbf{q})=\langle u_\mathrm{HF}(\mathbf{k_1+q})|P_l|u_\mathrm{HF}(\mathbf{k_1})\rangle$ is the form factor of the Bloch wavefunction of the HF band on layer $l$. Upon a particle-hole transformation $\tilde{c}_0(\mathbf{k})\rightarrow \tilde{h}_0^\dagger(\mathbf{k})$, we obtain:
\begin{equation}\label{eqED}
\begin{split}
    H=&-\sum_\mathbf{k}\left( \epsilon(\mathbf{k})+2\sum_{\mathbf{k^\prime}}\left(V(\mathbf{k},\mathbf{k^\prime},\mathbf{0})-V(\mathbf{k},\mathbf{k^\prime},\mathbf{k^\prime-k})\right)\right)\tilde{h}_0^\dagger(\mathbf{k})\tilde{h}_0(\mathbf{k})\\
    &+\sum_{\mathbf{k_1},\mathbf{k_2},\mathbf{q}}V(\mathbf{k_1},\mathbf{k_2},\mathbf{q})\tilde{h}_0^\dagger(\mathbf{k_1+q})\tilde{h}_0^\dagger(\mathbf{k_2-q})\tilde{h}_0(\mathbf{k_2})\tilde{h}_0(\mathbf{k_1})+E_0.
\end{split}
\end{equation}
Here $\epsilon(\mathbf{k})-\sum_{\mathbf{k^\prime}}\left(V(\mathbf{k},\mathbf{k^\prime},\mathbf{0})-V(\mathbf{k},\mathbf{k^\prime},\mathbf{k^\prime-k})\right)$ is the HF dispersion. Therefore, we use the hole picture relative to the $\nu=1$ state, applying the dispersion and the form factor of the HF band to perform the ED calculation.

The energy constant $E_0$ is the energy of the $\nu=1$ insulator, it is evaluated as:
\begin{equation}\label{eqE0}
    E_0=\sum_\mathbf{k} \left(\epsilon(\mathbf{k}) + \sum_{\mathbf{k^\prime}}\left(V(\mathbf{k},\mathbf{k^\prime},0)-V(\mathbf{k},\mathbf{k^\prime},\mathbf{k^\prime-k})\right)\right).
\end{equation}

In the ED calculation, we define our many-body Hamiltonian in the hole picture relative to the $\nu=1$ spin-valley polarized Chern insulator. Note that the naive way of doing the calculation in the electron picture has the problem of double counting.  This means that the $\nu=\frac{2}{3}$ filling should be thought as $1/3$ filling in terms of holes respect to the  Chern insulator.  Because the $\nu=1$ parent state is spin-valley polarized, the hole creation operator of the other spin or valley species is meaningless. So we can only access spin-valley polarized FCI states in this framework.

\section{Demonstration of convergence in Hartree Fock and ED calculations}
\begin{figure}[H]
\centering
    \includegraphics[width=0.8\linewidth]{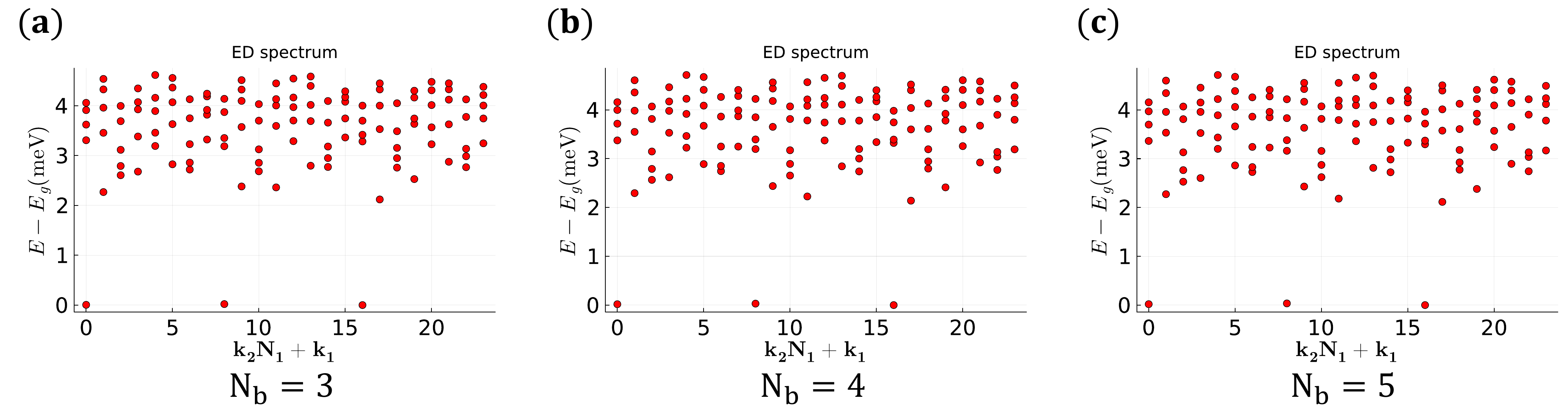}
    \caption{ED calculation of the $5$-layer graphene/hBN system at $\theta=0.77^\circ$($a_M = 11.36\mathrm{nm}$), $\epsilon=6$, $D=-160\mathrm{meV}$. The system size is fixed to be $4\times6$. We vary $N_b=3,4,5$ to demonstrate that the results have converged already at $N_b=4$.}
    \label{fig:changeNb}
\end{figure}

We keep $N_b$ number of conduction bands in the MBZ to perform the Hartree Fock calculation. In the main text we use $N_b=5$. To confirm that the calculation converges with $N_b$, we also do HF and ED calculations with $N_b=3,4$. The result is shown in Fig.~\ref{fig:changeNb}. One can see that the many body gap and ED spectrum at $\nu=2/3$ does not vary for $N_b=3,4,5$. This proves that $N_b=5$ is sufficient.

\begin{figure}[H]
\centering
    \includegraphics[width=0.8\linewidth]{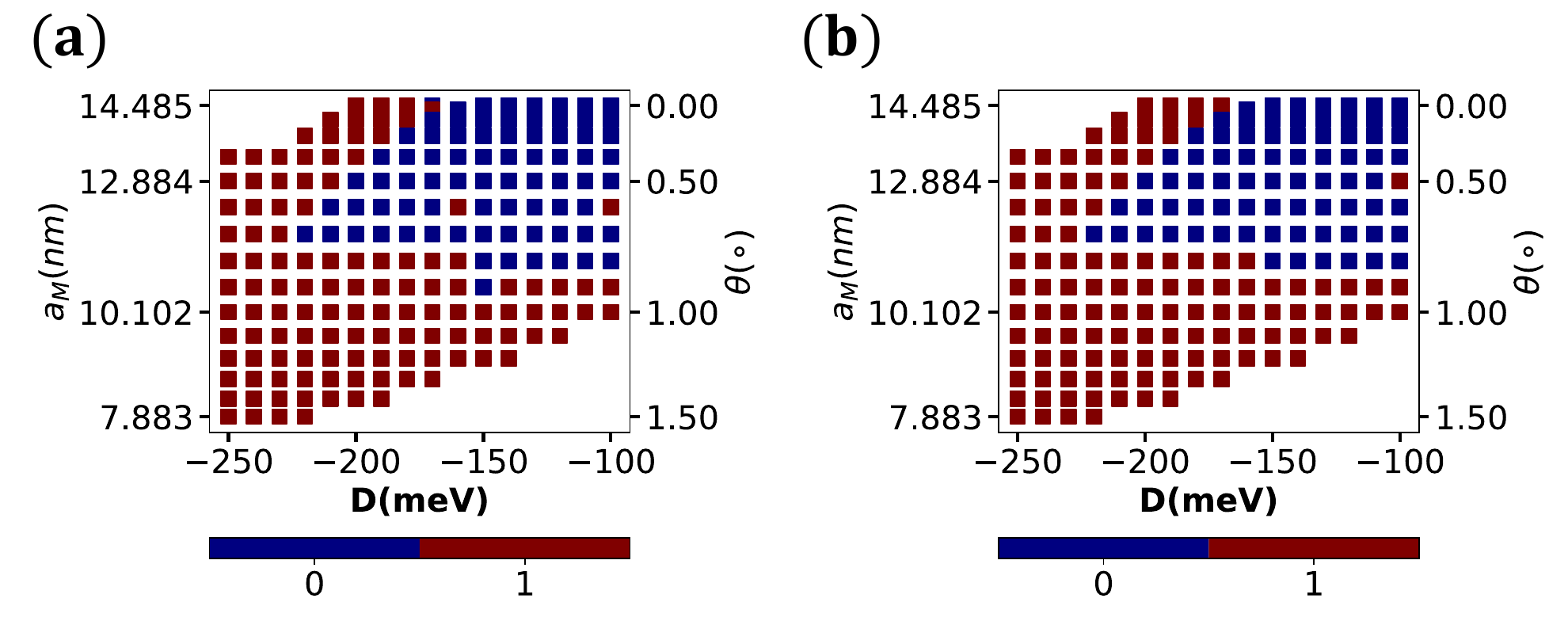}
    \caption{HF calculation results of the 5-layer graphene/hBN system at $\epsilon=6, V_M=0$ with different system sizes. (a) System size of $12\times12$. (b) System size of $12\times 18$. The color indicates the lowest band's Chern number after HF. The blank area corresponds to the metal phase. }
    \label{fig:different_size_cdw}
\end{figure}

Moreover, we also check the convergence on the system size in the Hartree Fock calculation. We show the phase diagram of the Chern number and the band gap in Fig.~\ref{fig:different_size_cdw}. The results are the same for HF calculation with system size $12\times 12$ and $12\times18$. 

\begin{figure}[H]
\centering

\includegraphics[width=0.8\linewidth]{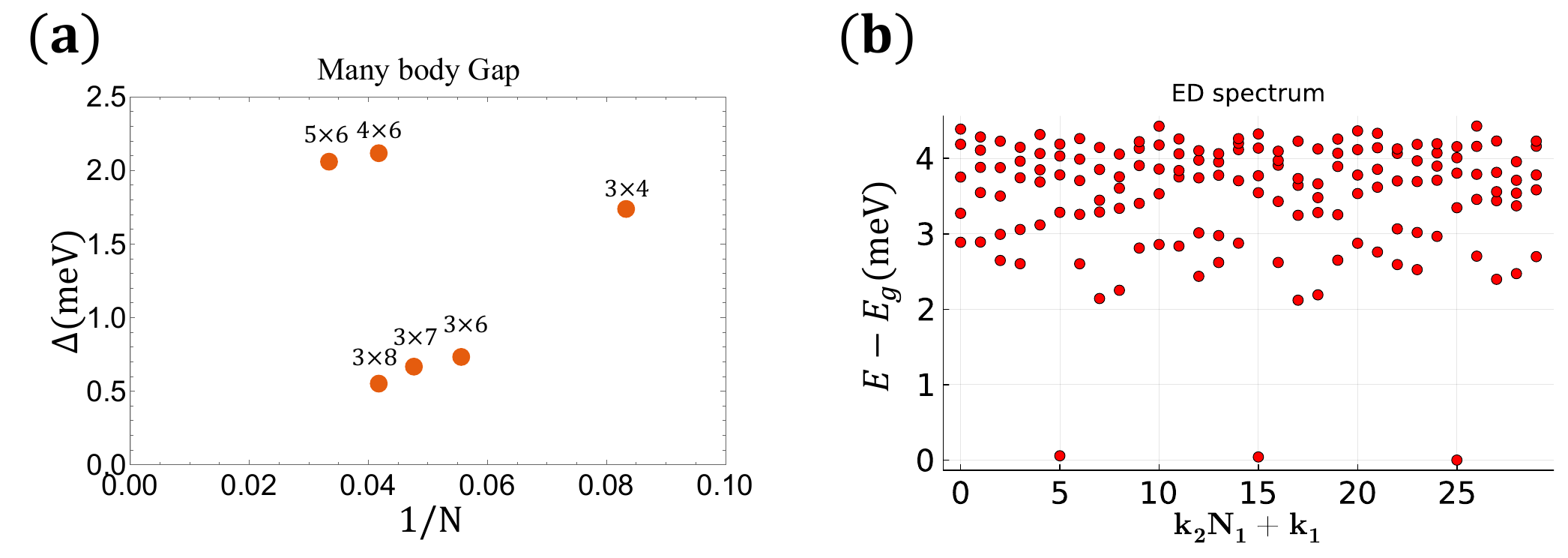}
\caption{(a) The many body gap for different system sizes at $\nu=\frac{2}{3}$. The many body gap is defined as the energy difference between the fourth and the third states in the ED calculation. (b) ED spectrum of system size $5\times6$.}
\label{fig:ED_finite_size_scaling}    
\end{figure}

In Fig.~\ref{fig:ED_finite_size_scaling}(a), we perform the ED calculation at $\nu=\frac{2}{3}$ in different system sizes. In the calculation we use $N_b=4$, $\theta=0.77^\circ$, $D=-160$meV and $V_M=1$.  Our result suggests that the gap would persist in the thermodynamic limit. In the $3\times L$ sequence, the gap decreases with $L$. But the many body gap becomes larger for $4\times 6$ and $5\times 6$. The ED spectrum of system size $5\times6$ is shown in Fig.~\ref{fig:ED_finite_size_scaling}(b). The same dependence on system size was found in ED calculation of twisted MoTe2 in Ref.\onlinecite{li2021spontaneous}.

\section{ED results at other fillings}
\begin{figure}[H]
\centering
    \includegraphics[width=0.6\linewidth]{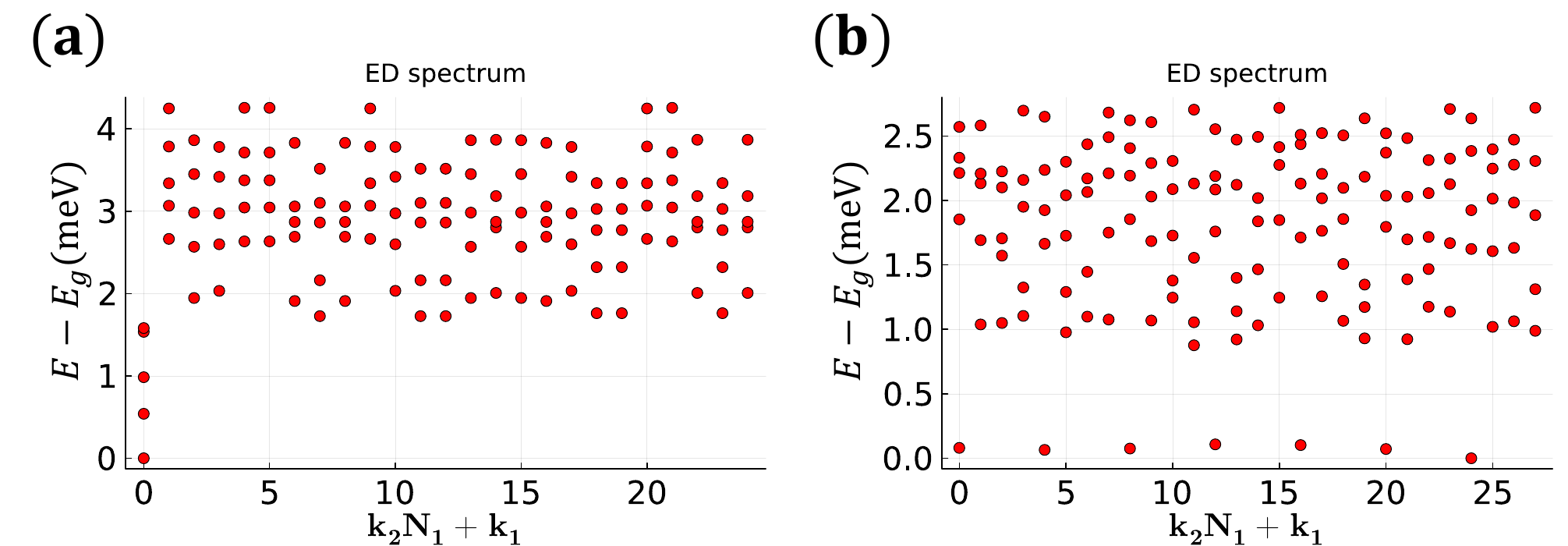}
    \caption{ED calculation of the 5-layer graphene/hBN system at $\theta=0.77^\circ(a_M=11.36\mathrm{nm})$, $\epsilon=6$, $D=-160\mathrm{meV}$. (a) ED calculation at $\nu=\frac{2}{5}$ with a system size of $5\times5$. (Based on the HF calculation performed with system size $20\times20$.) (b) ED calculation at $\nu=\frac{3}{7}$ with a system size of $4\times7$. (Based on the HF calculation performed with system size $24\times28$.)}
    \label{fig:more_fillings}
\end{figure}

In Fig.~\ref{fig:more_fillings}, we show the ED spectrum of $\nu=\frac{2}{5},\frac{3}{7}$, indicating FCI states at these fillings.

\subsection{CFL at $1/2$ filling}

\begin{figure}[H]
\centering    
    \includegraphics[width=0.9\linewidth]{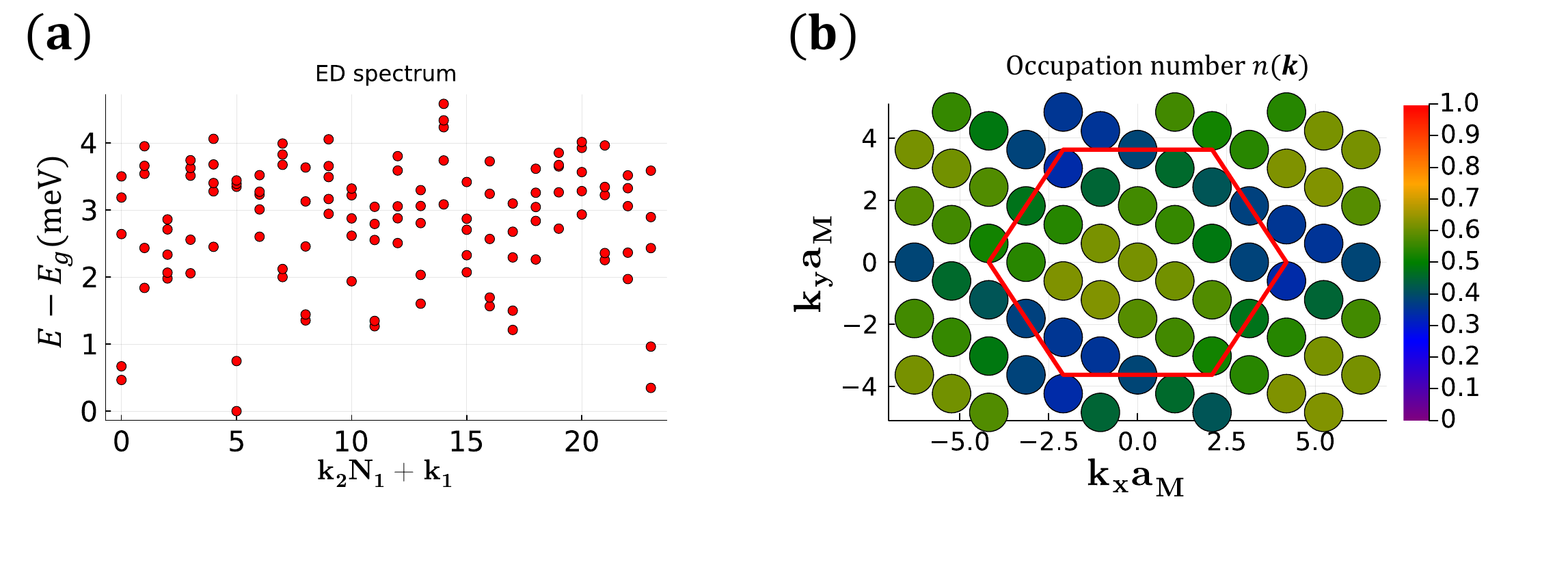}
    \caption{(a) ED calculation at $\nu=1/2$ filling with system size of $4\times6$ and the momentum distribution $n(\mathbf{k})=\langle c^\dagger(\mathbf{k})c(\mathbf{k})\rangle$. The calculations are done for $\theta=0.9^\circ$, $\epsilon=6$, $D=-160$ meV and we use $N_b=5$.}
    \label{fig:1/2filling}
\end{figure}

With parameters $\theta=0.9^\circ$($a_M=10.63\mathrm{nm}$), $\epsilon=6$, $D=-160\mathrm{meV}$, we perform an ED calculation at $\nu=1/2$ filling in system size of $4\times6$. The result is shown  in Fig.~\ref{fig:1/2filling}. We perform a HF calculation at a $24\times24$ system and extracts a $4\times6$ submesh to perform the ED calculation. As shown in Fig.~\ref{fig:1/2filling}, we do not see any many body gap in the ED spectrum, implying a gapless state. Meanwhile the momentum distribution $n(\mathbf{k})$ is relatively flat. Therefore the phase is consistent with the composite Fermi liquid (CFL) state.

\section{Improved ED calculation at fractional filling beyond the rigid band approximation}\label{secalpha}
The Hartree Fock approach provides the unitary transformation to minimize the total energy at $\nu=1$. At fractional filling $\nu<1$, the Coulomb interaction correction to the system may be weakened.

Our ED calculation at fractional filling $\nu$ assumes a rigid band from the HF calculation at $\nu=1$. The flat Chern band is generated from the mean field Hamiltonian $H_\mathrm{HF}$ driven by the interaction. But there is worry that the spontaneously formed crystal may be weakened by the doping and then the above ED calculation may be incorrect. To allow the band to deform with the doping, we introduce a variational parameter $\alpha\in (0,1]$. Then we use the first band of the following Hamiltonian to do the ED calculation:
\begin{equation}\label{eqalpha}
    H_{\alpha}=H_K+\alpha H_{\textrm{HF}},
\end{equation}
where $H_K=H_0+H_M$ as defined in the main text (we use $V_M=1$ here), $H_{\textrm{HF}}=H_{\textrm{Hartree}}+H_{\textrm{Fock}}$ is the mean field decomposition of the Coulomb interaction at $\nu=1$. Then for each parameter $\alpha$, the diagonalization of $H_{\alpha}$ provides a unitary transformation as: $\tilde{c}_{m;\alpha}^\dagger(\mathbf{k})=U_{mn;\alpha}(\mathbf{k})c_n^\dagger(\mathbf{k})$, where $c_n(\mathbf{k})$ is the electron operator in the band of the bare $H_K$. $\tilde{c}_{m;\alpha}(\mathbf{k})$ is the electron operator in the band of $H_\alpha$. $m,n$ is the band index.  After that, we perform the ED calculation following the same procedure as in Sec.~\ref{Section_ED_calculation}. We still use the hole operator $\tilde h_0(\mathbf k)=\tilde c_0^\dagger(\mathbf k)$ and use Eq.~\ref{eqED} to calculate the energy. Here the dispersion term and the interaction vertex depend on $\alpha$ because we use the Bloch wavefunction of $H_\alpha$.



We need to emphasize that here we are using a variational wavefunction approach. $H_\alpha$ is simply used to define the state $\ket{\psi_\alpha}$, the ground state of the ED of the original Hamiltonian projected to a subspace depending on $\alpha$. But the energy is still the expectation value of the original Hamiltonian.  Then we can optimize the variational parameter $\alpha$ by minimizing the energy $E(\alpha)=\braket{\psi_\alpha|H|\psi_\alpha}$. If the spontaneously formed crystal is destroyed by doping, we expect the optimal $\alpha$ to be close to $0$. 

We also define a minimal wavefunction overlap of the lowest band $H_{\alpha}$ with $H_{\alpha=1}$ as the fidelity:
\begin{equation}
    F_{\alpha}=\mathrm{min}_\mathbf{k}\{|\sum_n U^*_{0n;\alpha}(\mathbf{k})U_{0n;\alpha}(\mathbf{k})|:0<\alpha\le1\},
\end{equation}  
which characterizes the deformation of the band.

\begin{figure}[H]
\centering
    \includegraphics[width=0.6\linewidth]{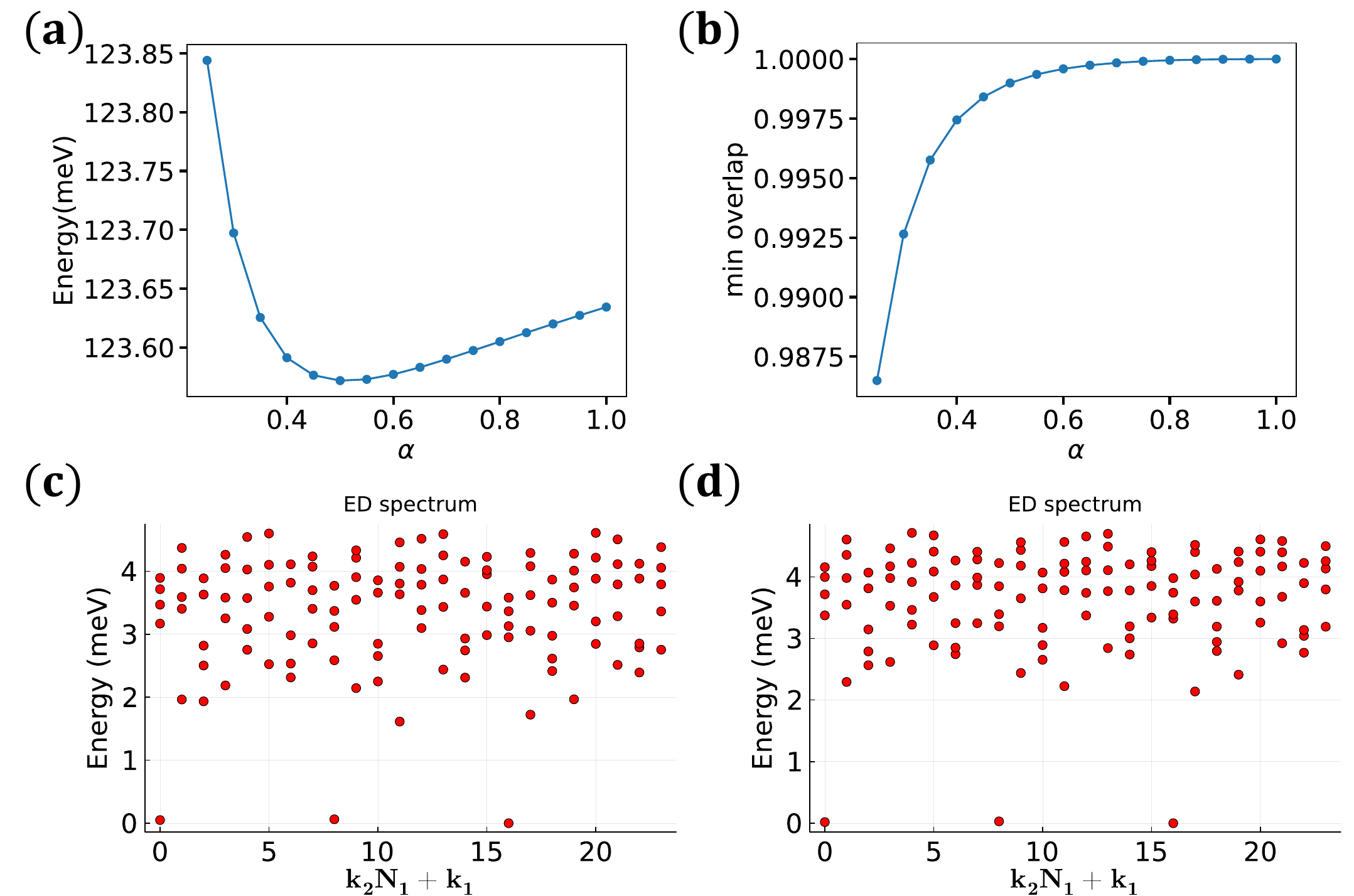}
    \caption{Calculation of the 5-layer graphene/hBN system at $\theta=0.77^\circ(a_M=11.36\mathrm{nm})$, $\epsilon=6$, $D=-160\mathrm{meV}$. The HF calculation is performed in the size of $12\times18$. The ED calculation is performed with a size of $4\times 6$. We keep $N_b=4$ conduction bands in our calculation. (a) Dependence of the energy per moir\'e unit cell on $\alpha$ at $\nu=\frac{2}{3}$. The energy minimum is at $\nu=0.5$. (b) Dependence of the fidelity $F_{\alpha}$ on $\alpha$. (c) ED spectrum at $\nu=\frac{2}{3}$ for $\alpha=1$. (d) ED spectrum at $\nu=\frac{2}{3}$ for $\alpha=0.5$.}
    \label{fig:change_alphaM}
\end{figure}

We perform the calculation at $\theta=0.77^\circ$ and $D=-160$ meV for the filling $\nu=\frac{2}{3}$. We find the optimal $\alpha$ to be $\alpha \approx 0.5$, as shown in Fig.~\ref{fig:change_alphaM} (a). We perform the ED calculation in the size of $4\times6$.  Meanwhile the fidelity has $F_{\alpha}>0.9$ for the whole range of $\alpha$, as shown in Fig.~\ref{fig:change_alphaM}(b). This is simply because that $H_\mathrm{HF}$ dominates over $H_K$ and the band deforms only a little.  As a result, the ED spectra with the optimal $\alpha\approx0.5$ and the rigid band with $\alpha=1$ at $\nu=\frac{2}{3}$ are basically the same, as shown in Fig.~\ref{fig:change_alphaM}(c)(d). Therefore we conclude that the rigid band approximation is justified and our previous ED calculations at fractional fillings are valid.
\section{Energy comparison between FCI and QAH crystal}
There are two competing states at fractional filling $\nu$: (I) An FQAH state with the crystal lattice fixed to the moir\'e lattice constant $a_M$; (II) An extended integer QAH state where the  Wigner crystal period $a_{\mathrm{crystal}}$  is spontaneously generated and scales with the filling $\nu$ as $a_{\mathrm{crystal}}=\frac{a_M}{\sqrt{\nu}}$. We calculate the case with $\nu=\frac{2}{3}$ in the main text. For (I), we perform HF calculation at a system size of $12\times18$, then choose a submesh of $4\times6$ for ED calculation by diagonalizing Eq.~\ref{eqED}. We tune $\alpha$ defined in Eq.~\ref{eqalpha} to get the minimal energy. For (II), we set $V_M=0$ first, then perform HF calculation at a system size of $12\times18$ with lattice constant $a_\mathrm{crystal}=\sqrt{\frac{3}{2}}a_M$. Consider the weak moir\'e potential as a first order perturbation, the incommensurance of $a_M$ and $a_\mathrm{crystal}$ results in $\langle H_M\rangle=0$.
\section{Visualization of the spontaneous generated crystal}

As discussed in the main text, the external moir\'e potential is only a tiny perturbation. We find interaction driven spontaneous formation of crystals. Here in Fig.~\ref{fig:density_profile}, we show the density distribution $\rho(\mathbf{r})$ at $\nu=1$ to visualize the crystal structure. $\rho(\mathbf{r})$ is defined as $\sum_{l}\langle\rho_l(\mathbf{q})\rangle e^{-\mathrm{i}\mathbf{q\cdot r}}$, $\rho_l(\mathbf{q})$ follows the same definition as in Eq.\ref{EqInteraction}. The $\rho(\mathbf{r})$ before doing HF calculation is shown in Fig.~\ref{fig:density_profile}(a). While $\epsilon=6,D=-160\mathrm{meV},\theta=0.77^\circ$, the solution with lowest energy has $C=1$, the corresponding $\rho(\mathbf{r})$ is shown in Fig.~\ref{fig:density_profile}(b). There are also solutions with $C=0$ with other parameters, for example while $\epsilon=6,D=-160\mathrm{meV},\theta=0.3^\circ$, the corresponding $\rho(\mathbf{r})$ is shown in Fig.~\ref{fig:density_profile}(c).

\begin{figure}[H]
\centering
    \includegraphics[width=0.8\linewidth]{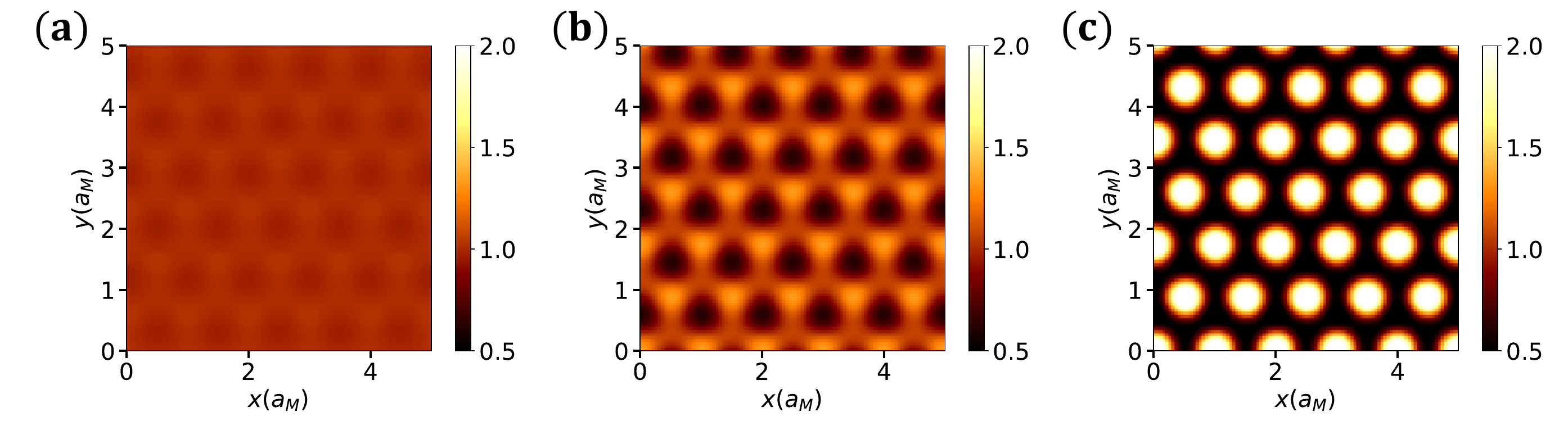}
    \caption{Numerical results of the 5-layer graphene/hBN system at $\epsilon=6$, $D=-160\mathrm{meV}$. The density distribution $\rho(\mathrm{r})$ at $\nu=1$: (a) $\theta=0.77^\circ$, without HF; (b) $\theta=0.77^\circ$, with HF, the lowest HF band has $C=1$; (c) $\theta=0.3^\circ$, with HF, the lowest HF band has $C=0$. }
    \label{fig:density_profile}
\end{figure}

\section{Results in other number of layers}
\begin{figure}[H]   
\centering
    \includegraphics[width=0.8\linewidth]{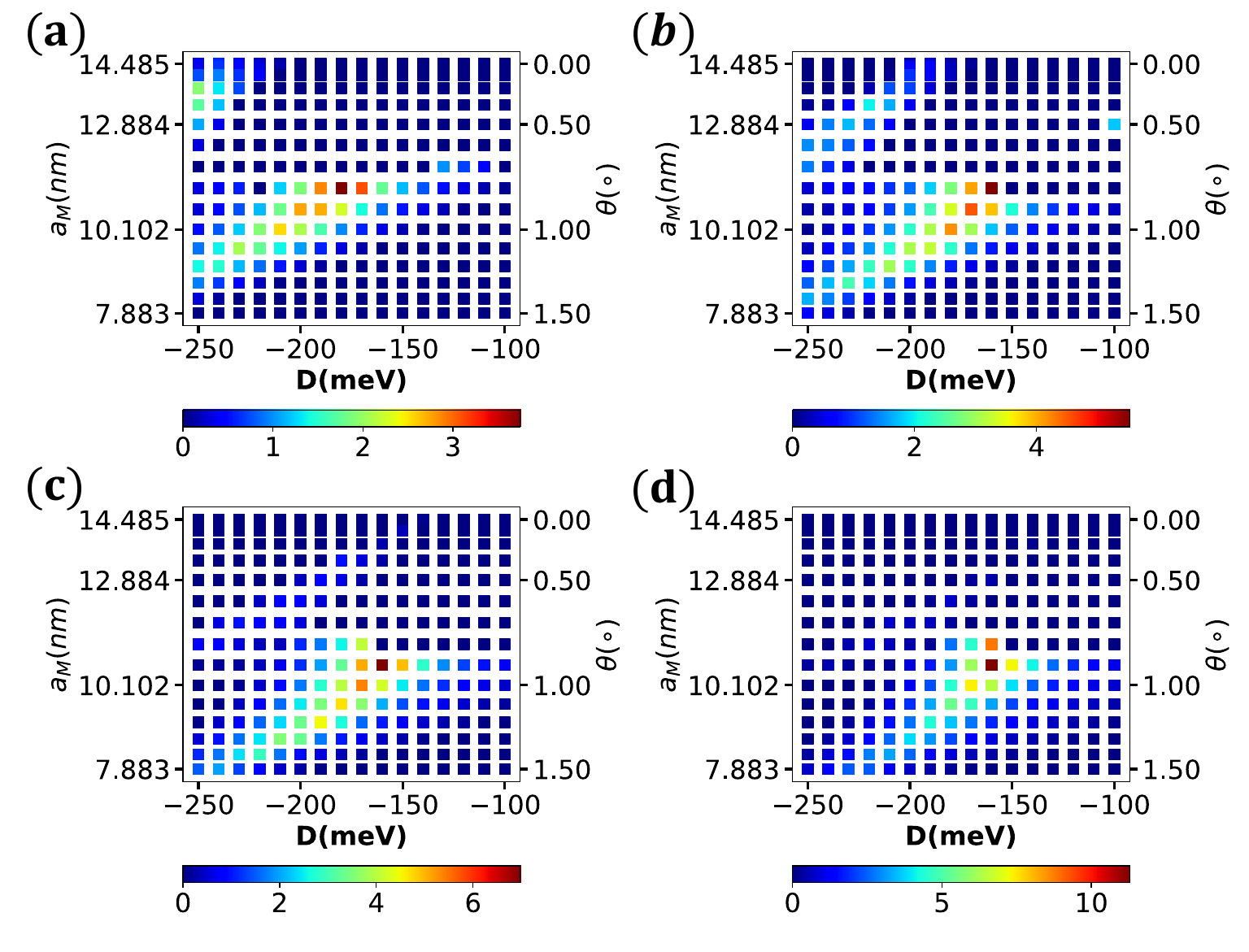}
    \caption{The HF calculation is performed by using $12\times 18$ points in MBZ. $\epsilon=6, V_M=0$. Dependence of $\frac{|C|\Delta}{W}$ on $D$ and $a_M$ of the $n$-layer graphene/hBN system. (a) $4-$layer. (b) $5-$layer. (c) $6-$layer. (d) $7-$layer. In (c)(d), we use $N_b=4$ for HF calculation. The maximum value of $\frac{|C|\Delta}{W}$ becomes larger as the number of layer increases.}
    \label{fig:different_layers}
\end{figure}

Besides the 5-layer graphene/hBN system, we also find that the same $C=1$  QAH crystal at $\nu=1$ exists also for $n_{\mathrm{layer}}=4,6,7$ as shown in Fig.~\ref{fig:different_layers}. Actually the indicator $\frac{|C| \Delta}{W}$ at the optimal region becomes larger when increasing $n_{\mathrm{layer}}$, suggesting that 6-layer and 7-layer system may host even more robust QAH and FQAH states than the 5-layer system already studied in the current experiment.

\section{The moir\'e potential from TBG or MG}
\begin{figure}[H]
\centering
    \includegraphics[scale=0.6]{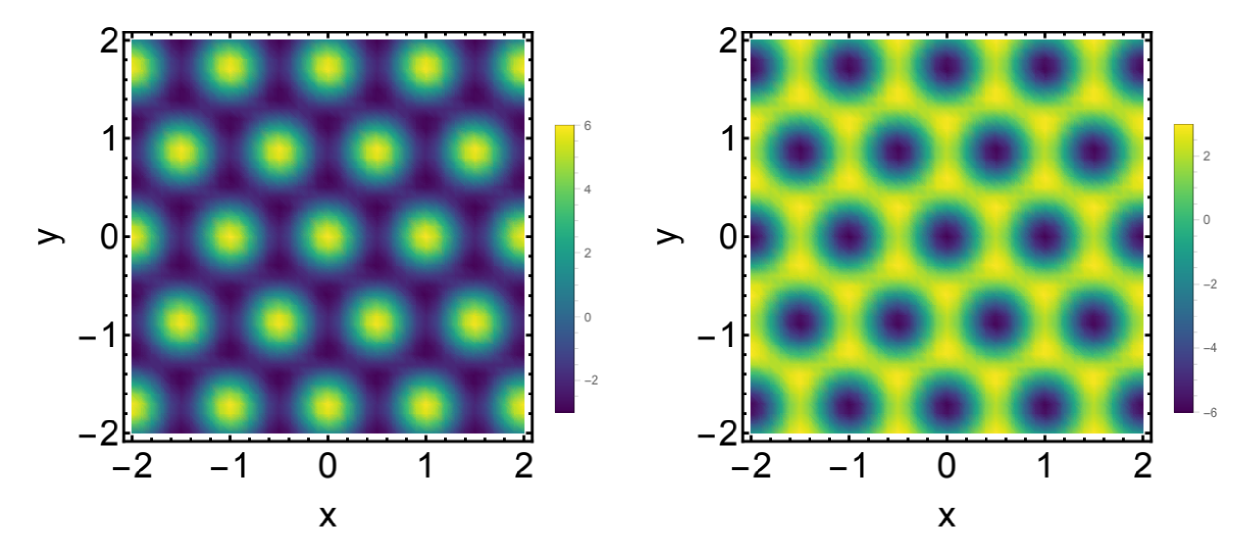}
    \caption{The moir\'e potential in real space, with $x$ and $y$ in units $a_M$. The left(right) hand side corresponds to different fillings of TBG. At $\nu_{TBG}=4$ (left hand side), the potential is positive, which makes the graphene multilayers favor honeycomb lattice, while $\nu_{TBG}=-4$ (right hand side), the potential is negative, which makes the graphene multilayers favor triangular lattice.}
    \label{real_space_VM}
\end{figure}

 We consider a different way to generate the moir\'e potential in the $n$-layer graphene through Coulomb interaction. We separate it with twisted bilayer graphene (TBG) or monolayer graphene (aligned with the middle hBN) through a thin hBN with thickness $d$. The coulomb interaction between TBG (or monolayer graphene) and the graphene layers is:
 \begin{align}
     H_V=\frac{1}{A}\sum_{\mathbf{q},l}V_l({\bf q})\rho_{t,l}({\bf q})\rho_b(-{\bf q}),
 \end{align}
 where $\rho_b({\bf q})=\sum_{\bf k}f_b^\dag({\bf k}+{\bf q}) f_b({\bf k})$ and $\rho_{t,l}(-{\bf q})=\sum_{\bf k}f_{t,l}^\dag({\bf k}-{\bf q}) f_{t,l}({\bf k})$ correspond to the density in the twisted bilayer graphene, and the density in each layer in the multilayer graphene respectively. $A=\frac{4\pi^2N_{\mathrm{cell}}}{|\bf{G}_1\times{\bf G}_2|}$ is the sample area. $V_l({\bf q})=\frac{e^2}{2\epsilon\epsilon_0|{\bf q}|}e^{-|{\bf q}|(d+ld_\mathrm{layer})}$ is the screened Coulomb potential in the materials, and $\epsilon$ is the dielectric constant, $d_\mathrm{layer}$ is the interlayer distance of the graphene multilayers. In the discretized  momentum space, the interaction $H_V$ becomes:
 \begin{align}
     H_V&=\frac{2}{\sqrt{3}}\frac{1}{N_{\mathrm{cell}}}\sum_{{\mathbf{q},l}}\frac{e^2}{2\epsilon\epsilon_0a_M}\frac{1}{|{\bf q}|a_M}e^{-|{\bf q}|(d+ld_\mathrm{layer})}\rho_b({\bf q})\rho_{t,l}(-{\bf q})\notag\\
     &=\frac{2}{\sqrt{3}}\frac{1}{N_{\mathrm{cell}}}\sum_{{\mathbf{q},l}}\frac{e^2}{2\epsilon\epsilon_0a_M}\frac{1}{|{\bf q}|a_M}e^{-|{\bf q}|(d+ld_\mathrm{layer})}\rho_b({\bf q})\sum_{{\bf k}_2}f_{t,l}^\dag({\bf k}_2-{\bf q}) f_{t,l}({\bf k}_2).
 \end{align}
 The density in the twisted bilayer graphene can be approximated as:
 \begin{align}
     \rho_b({\bf q})=\sum_{\bf k}c_b^\dag({\bf k}+{\bf q})\Lambda_b({\bf k},{\bf q})c_b({\bf k}),
 \end{align}
 where $\Lambda_b({\bf k},{\bf q})=\langle u_b(\mathbf{k+q})|u_b(\mathbf{k})\rangle$ is the form factor, $c_b(\mathbf{k})$ is the annihilation operator of the electrons in the conduction band. The potential generated to the $n$-layer graphene is:
 \begin{align}
     V_l({\bf G}_M)=\frac{2}{\sqrt{3}}\frac{e^2}{2\epsilon\epsilon_0a_M}e^{-G_M{(d+ld_\mathrm{layer})}}\frac{1}{G_Ma_M}\langle\rho_b({\bf G}_M)\rangle,
 \end{align}
where $d$ is also the distance between the TBG (or MG) and the closest  layer of multilayer graphene. Finally, we find the potential in the layer $l=0,1,...,n_\mathrm{layer}-1$ is $V_l({\bf G}_M)=V_0({\bf G}_M)e^{-G_Mld_\mathrm{layer}}$, where $V_0({\bf G}_M)$ is the strength of the potential in the first graphene layer and it depends on the twisted angle and the distance between the hBN and the graphene layer. Fourier transforming the term $\sum_{{\bf k}}\sum_{{\bf G}_i}V_M({\bf G}_i)c^\dag_{{\bf k}+{\bf G}_i}c_{\bf k}$ into real space, we have $\sum_{i}V(R_i)c^\dag(R_i)c(R_i)$, with $V({R_i})=\sum_{{\bf G}_i}V({\bf G}_i)e^{\mathbbm{i}{\bf G}_i\cdot {\bf R}_i}$. The real space potential is shown in Fig.~\ref{real_space_VM}, from which we can see for positive $V_M$ the system prefers honeycomb lattice, while for negative $V_M$ it prefers triangular lattice.

\begin{figure}[H]    
\centering
    \includegraphics[scale=0.4]{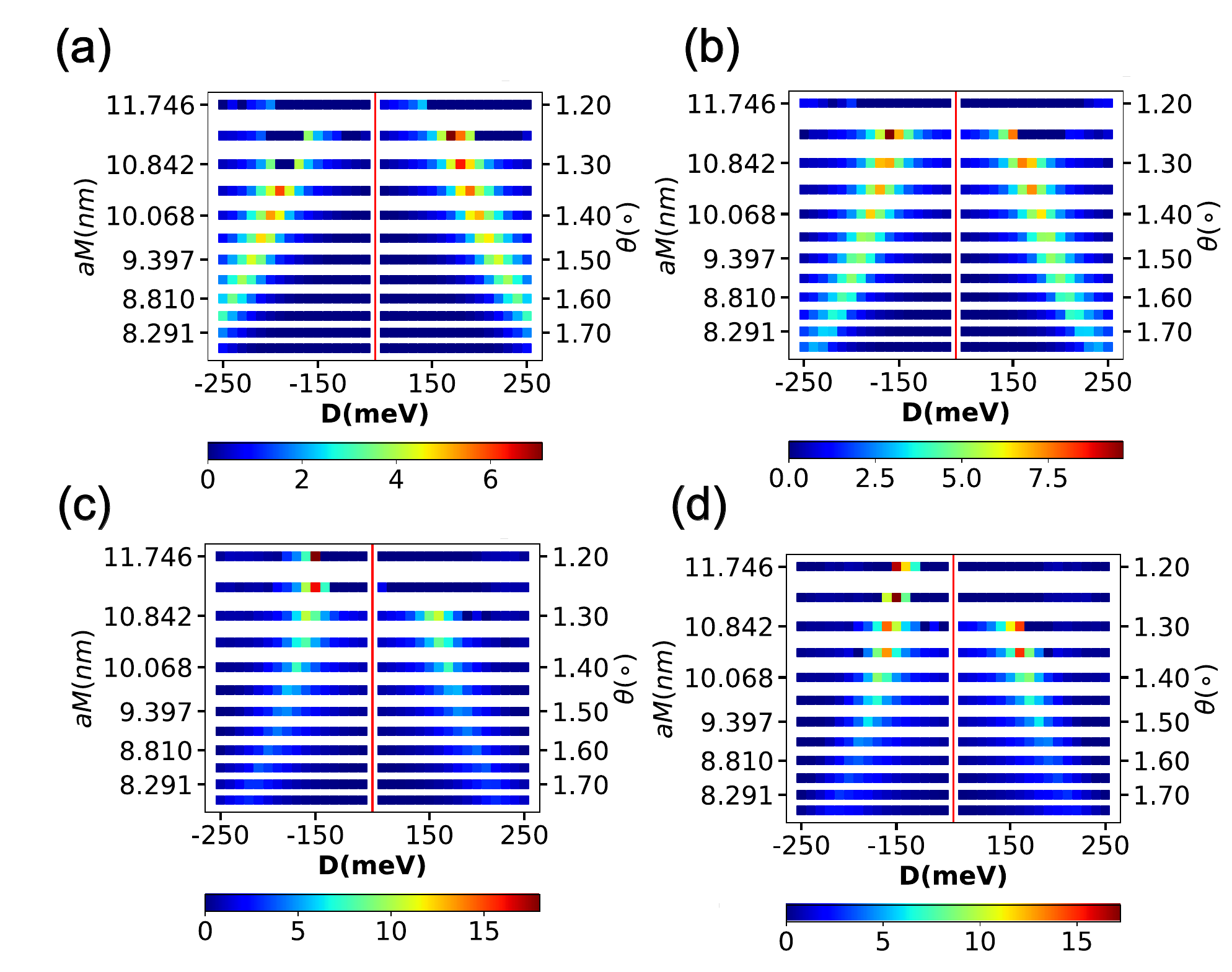}
    \caption{Dependence of $\frac{|C|\Delta}{W}$ on $D$ and $a_M$ of the $n-$layer graphene/hBN/TBG set up for (a) 4-layer, (b) 5-layer, (c) 6-layer, (d) 7-layer, $|C|$, $\Delta$ and $W$ follow the same definition as in Fig.2 in the main text. Here we use $\epsilon=6, N_b=4$, and the distance between the TBG and $n-$layer graphene is $d=5$nm.}
    \label{fig:TBG_layer}
\end{figure}

We perform the Hartree Fock calculation and draw a phase diagram for the case with TBG in Fig.~\ref{fig:TBG_layer}. The result is quite similar to Fig. 3 in the main text. We find that a narrow $C=1$ Chern band is possible for 4-layer, 5-layer, 6-layer and 7-layer graphene system with $a_M$ in a range smaller than 11nm. Now the $D>0$ side also  hosts a narrow Chern band. 

\begin{figure}[H]
\centering
    \includegraphics[scale=0.4]{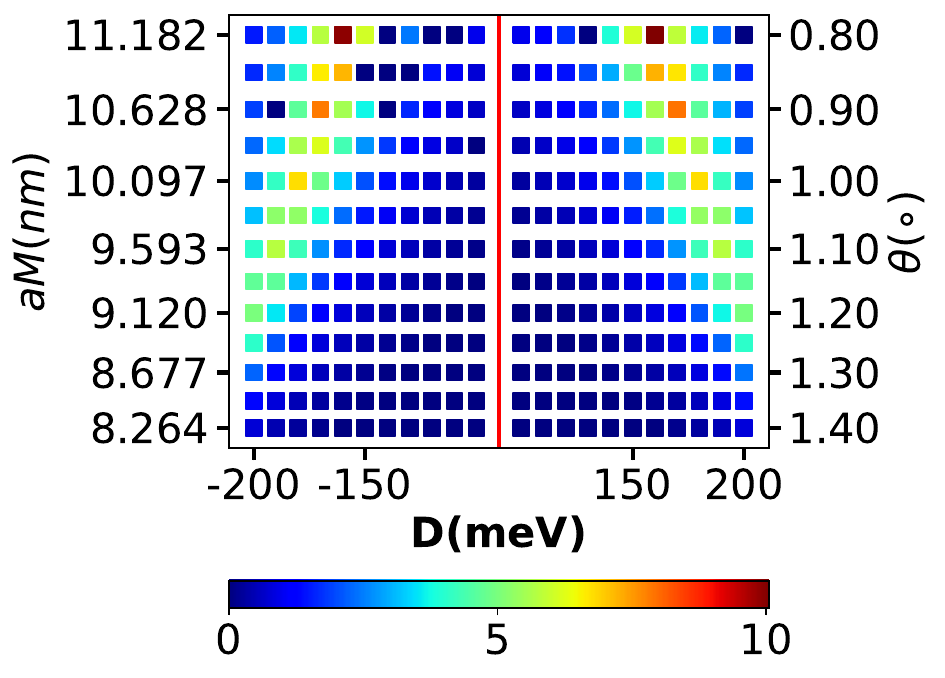}
    \caption{$\frac{|C|\Delta}{W}$ of 5-layer graphene/hBN/MG system. $a_M$ is controlled by the twist angle of the MG and the middle hBN. We use $\epsilon=6$, $d=5\mathrm{nm}$.}
    \label{fig:MLG}
\end{figure}

We can also replace TBG with a monolayer graphene (MG) aligned with the middle hBN. The phase diagram is shown in Fig.~\ref{fig:MLG} . At $d=1\mathrm{nm}$, for $\epsilon=6$, the moir\'e potentials $V_0(G_M)$ in the $n$-layer graphene/hBN/TBG and $n$-layer graphene/hBN/MG are, $17.9$meV and $0.84$meV, respectively. In our Hartree Fock calculation we can get the QAH Wigner crystal in both cases as the moir\'e potential does not make any difference. But it is known that the Hartree Fock calculation may overestimate the strength of an insulating state, we leave to future experiment to see whether the weak moir\'e potential from the $n$-layer graphene/hBN/MG system is enough to stabilize the QAH crystal and FQAH phases.

\begin{figure}[H]
\centering
    \includegraphics[scale=0.4]{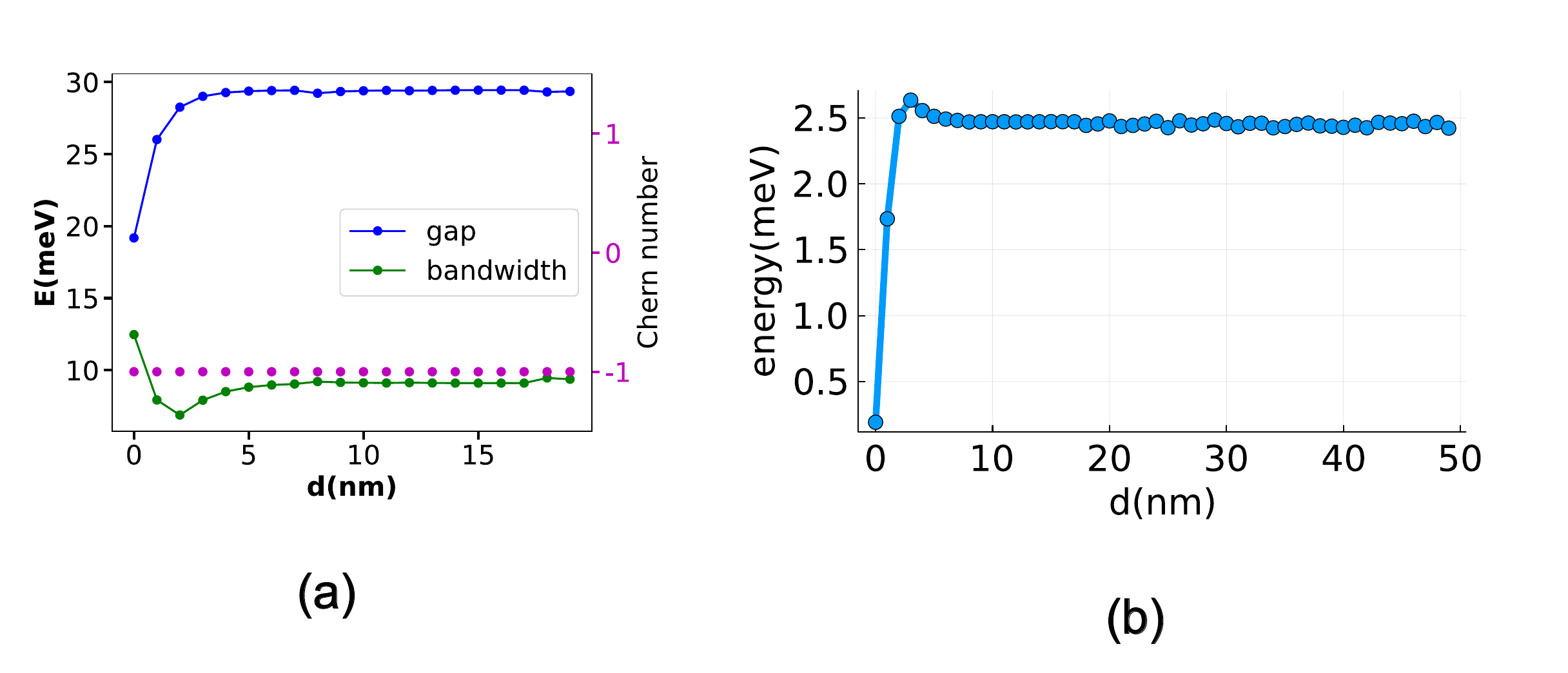}
    \caption{(a) The band gap, the band width and Chern number for the 5-layer graphene/hBN/TBG system at filling $\nu=1$ as a function of the distance $d$ between the TBG and $5-$layer graphene in the set up of Fig.4 in the main text. Here we use $D=180$meV. (b) The many body gap for $D=180$meV, at $\nu=\frac{2}{3}$ with system size $4\times6$. We use twisted angle $\theta=1.5^\circ (a_M\approx9.4\mathrm{nm})$, and dielectric constant $\epsilon=6$, $N_b=3$. One can see that FQAH is possible now also at $D>0$ side and survives to the $d\rightarrow+\infty$ limit.}
    \label{fig:TBG_change_d}
\end{figure}

In this new setup we can control the moir\'e potential by tuning the distance $d$. In Fig.~\ref{fig:TBG_change_d}, we  find that QAH-Wigner crystal phase, together with the FQAH phase, survive to the $d\rightarrow +\infty$ limit also for $D>0$, consistent with our conclusion that the external moir\'e potential is not essential for the Chern band physics.

\begin{figure}[htbp]    
   \includegraphics[width=0.7\linewidth]{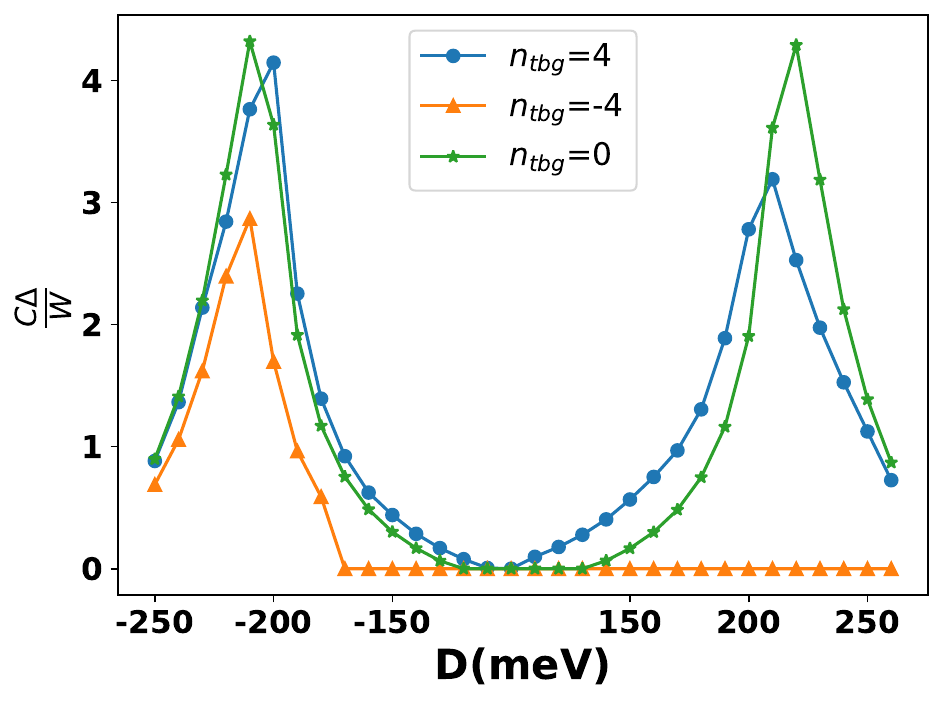}
   \caption{Displacement field dependence of $\frac{C\Delta}{W}$ for 5-layer graphene/hBN/TBG set up at $d=1\mathrm{nm}$, $\theta=1.6^\circ$ and $\epsilon=6$.}
   \label{fig:line_cut}
\end{figure}

\begin{figure}[htbp]
\centering
    \includegraphics[scale=0.4]{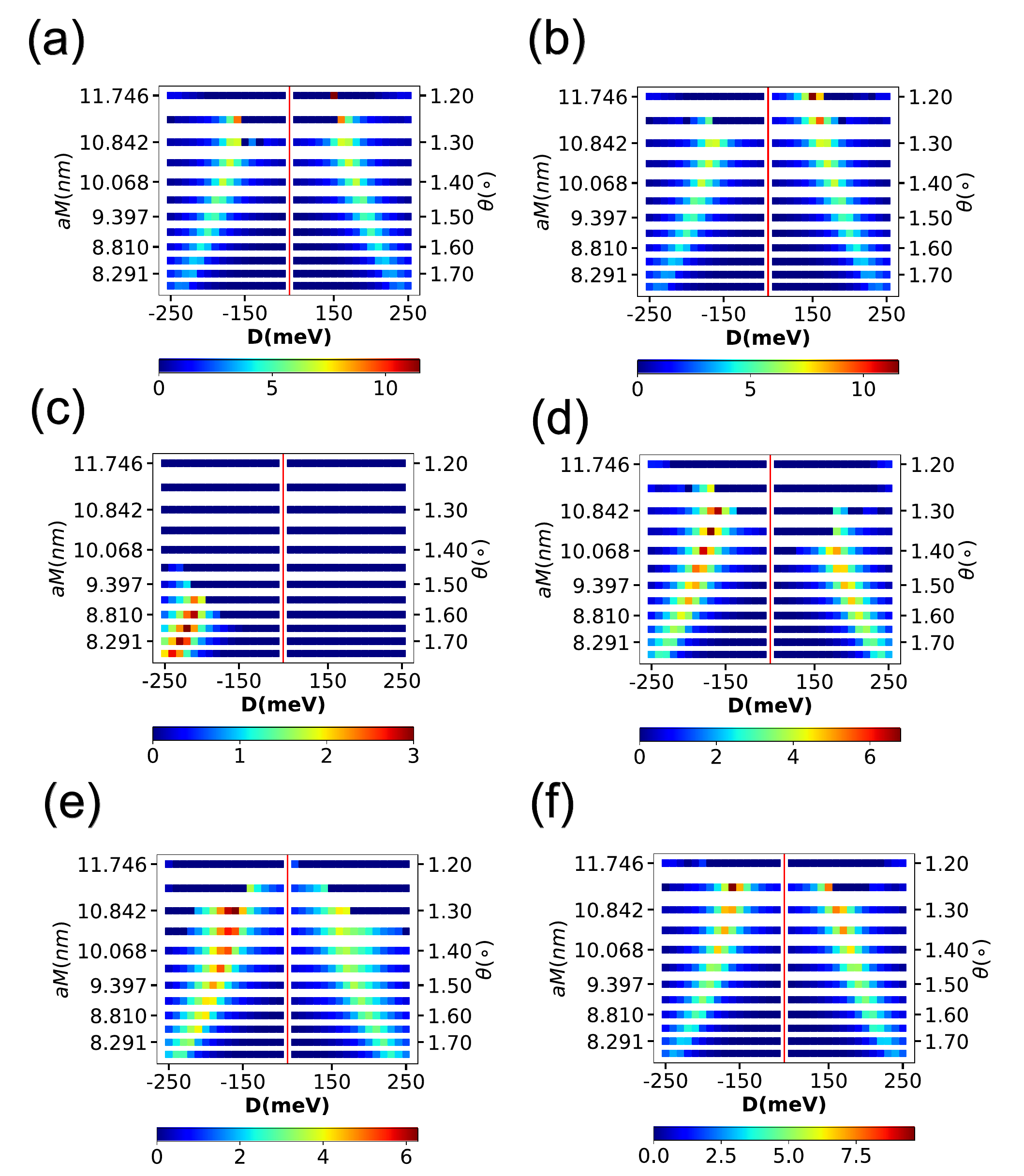}
    \caption{(a) (b) the phase diagram of 5-layer graphene for $\nu_{\mathrm{TBG}}=0$ and $d=1nm$ and $d=5nm$, respectively. (c) (d) the phase diagram of 5-layer graphene for $\nu_{\mathrm{TBG}}=-4$ and $d=1nm$ and $d=5nm$, respectively. (e) (f) the phase diagram of 5-layer graphene for $\nu_{\mathrm{TBG}}=4$ and $d=1nm$ and $d=5nm$, respectively.}
    \label{fig:phase_diagram_tune_tbg}
\end{figure}

In Fig.~\ref{fig:phase_diagram_tune_tbg}, and Fig.~\ref{fig:line_cut}, we show line cut of phase diagram and the whole phase diagrams at $\theta=1.6^\circ$. We can find the FCI phase is suppressed at $\nu_{TBG}=-4$ compared to $\nu_{TBG}=0,4$ with $d=1$ nm. At $\nu_{TBG}=-4$, the induced moir\'e superlattice is triangular and  disfavor the QAH crystal which needs a honeycomb structure. When $d=5$ nm, the induced moir\'e potential is weak, then the Hartree Fock results do not depend much on $\nu_{TBG}$ anymore.

\end{document}